\documentclass[12pt]{article}
\usepackage{graphics,longtable,amssymb}
\newcommand{\ba}{\begin{equation}}
\newcommand{\ea}{\end{equation}}
\newcommand{\ZWEI}[4]{\left \{ \begin{array}{ll} #1 & #2 \\ & \\ #3 & #4
\end{array} \right.}

\newcommand{\X}{\mbox{\boldmath $\displaystyle X$}}

\newcommand{\z}{\mbox{\boldmath $z$}}

\newcommand{\dx}{\mbox{\boldmath $d$}}

\newcommand{\ux}{\mbox{\boldmath $u$}}

\newcommand{\UGR}{\mbox{\boldmath $\displaystyle U$}}

\newcommand{\xalpha} {\mbox{\boldmath $\alpha$}}
\newcommand{\xbeta} {\mbox{\boldmath $\beta$}}

\newenvironment{fett}[1]{
 \begin{list}{}{
  \settowidth{\labelwidth}{\bf #1}
 \setlength{\leftmargin}{\labelwidth}
   \labelsep4mm
  \addtolength{\leftmargin}{\labelsep}
  
             }}{\end{list}}

\newcommand{\EINS}{1\!\mbox{I}}

\newcommand{\NAT} {\mathbb{N}}
\newcommand{\NATO} {\mathbb{N} _{0}}
\newcommand{\REL} {\mathbb{R}}

\newcounter{BUMACOUNT}[subsection]
\newcounter{BUMCOUNT}[section]
\newenvironment{bum}[1]
{ \refstepcounter{BUMCOUNT}
 \noindent {\bf \thesection.\arabic{BUMCOUNT}} \ \  {\bf #1}. \quad }
\newcommand{\be}{\begin{equation}}
\newcommand{\ee}{\end{equation}}
\newcommand{\wh}{\widehat}

\title
{\Large {\bf A Bivariate Dead Band Process Adjustment Policy
}}
\author{
{\small Enrique Del Castillo}\\
{\small Department of Industrial and Manufacturing Engineering and Department of Statistics,}\\
{\small The Pennsylvania State University,}\\
{\small University Park, PA 16802, USA}\\[3mm]
{\small and}\\[5mm]
{\small Rainer G\"{o}b}\\
{\small Institut f\"{u}r Angewandte Mathematik und Statistik, Universit\"{a}t
W\"{u}rzburg,}\\
{\small Sanderring 2, D-97070 W\"{u}rzburg, Germany}}\vspace{0.6cm}

\date{\small December 2019 }

\begin{document}

\maketitle
\begin{abstract}
A bivariate extension to Box and Jenkins (1963) feedback adjustment problem is presented in this paper. The model balances the fixed cost of making an adjustment, which is assumed independent of the magnitude of the adjustments, with the cost of running the process off-target, which is assumed quadratic. It is also assumed that two controllable factors are available to compensate for the deviations from target of two responses in the presence of a bivariate IMA(1,1) disturbance. The optimal policy has the form of a ``dead band", in which adjustments are justified only when the predicted process responses exceed some boundary in $\mathbb{R}^2$. This boundary indicates when the responses are predicted to be far enough from their targets that an additional adjustment or intervention in the process is justified.  Although originally developed to control a machine tool, dead band control policies have application in other areas. For example, they could be used to control a disease through the application of a drug to a patient depending on the level of a substance in the body (e.g., diabetes control). This paper presents analytical formulae for the computation of the loss function that combines off-target and adjustment costs per time unit.  Expressions are derived for the average adjustment interval and for the scaled mean square deviations from target. The minimization of the loss function and the practical use of the resulting dead band adjustment strategy is illustrated with an application to a  semiconductor manufacturing process.\\
\end{abstract}

%\vspace{10mm}

\section{Introduction.}
\label{Introduction}

In a landmark paper, Box and Jenkins (1963) contrasted adjustment policies for a ``chemical" process with those of a ``machine tool" process. The latter kind of process usually involves a large adjustment cost that is independent of the magnitude of the adjustments, in contrast to the former where off-target costs typically dominate. Assuming quadratic off-target costs, Box and Jenkins showed that the sum of off-target and fixed adjustment costs per time unit is minimized by a policy that has the form of what we will refer to in this paper as a {\em dead band adjustment policy}. In the univariate version of this type of policy, the process is not adjusted as long as the one step ahead  minimum mean square error (MMSE) forecast, if no adjustment is made, falls inside two ``control lines" placed symmetrically around the process target that define a band or region (the dead band) within which the process is let uncontrolled. The process if uncontrolled is assumed to drift off-target according to an IMA(1,1) process. The optimal policy resembles a Shewhart control chart applied to the forecasts, but the width of the control (or adjustment) limits is based on balancing the costs of running the process off-target and adjusting the process. For a description of the origins of this type of control problem and its relation to other types of process control problems, see Del Castillo (2002, 2006) and Woodall and Del Castillo (2014).

%{\em A dead band control policy has applicability in areas outside of process control of %machines. For example, it has the same form as certain drug delivery policies where a %drug is supplied to the patient depending on the level of one or more substances in the %body (e.g., glucose monitoring and insulin supply in intensive diabetes control). In the %present paper, we extend the Box-Jenkins univariate dead band model to the case there are %two responses of interest, possibly cross-correlated, and there are two controllable %factors available to adjusting the process.}

By far, the interest in dead band control exceeds  the originally considered machine tool
problem. Similar problems exist in other areas, e.g., in biosciences, business administration, or
financial engineering.
For instance, a dead band control policy
has the same form as certain drug delivery policies where a drug is supplied to the patient depending on the level of one or more substances in the body (e.g., glucose monitoring and insulin supply in intensive diabetes control, see, e.g., Magni et al. (2009)).
Another  area of interest is the control of cash flows, e.g., investment flows, or transfers between departments or different branches of a corporation. In the latter cases, adjustments often require relatively expensive interventions into the administration or organizational structure, i.e.,  adjustment costs are high.

Properties of univariate dead band adjustment policies have been studied by other authors. Crowder (1992) solves Box and Jenkins's univariate machine tool problem using dynamic programming techniques when there is a finite number of periods in the planning horizon for the process. This is in contrast to Box and Jenkins (1963), who use a renewal reward process to minimize the long-run average cost per time unit.
Crowder shows how the optimal dead band control limits funnel out as the end of the production run approaches, when a renewal of the process is assumed to occur. For the initial periods, the dead band limit width approaches the long-run solution obtained by Box-Jenkins as the planning horizon increases.
Box and Kramer (1992) added a sampling cost component to the univariate machine tool model and discuss dead band policies when samples are not taken at every period. Jensen and Varderman (1993) studied the finite-horizon model in Crowder (1992) but considered the possibility that adjustment errors occur when setting the controllable factor. They show that even when there is no fixed adjustment cost (only quadratic off-target costs are present), a dead band-like policy is the optimal policy in the presence of adjustment errors. Srivastava and Wu (1991) consider the machine tool problem
under the presence of inspection costs which
were not included in the original model by Box and Jenkins.
A related type of control problem where an optimal deadband policy results are setup adjustment problems where, under i.i.d noise (i.e., no drift) there is a fixed adjustment cost to compensate for sudden upsets, including at starting up a process, see Zilong and Del Castillo (2006), and for a multivariate generalization Liu et al. (2013).

In the present paper, we extend the Box-Jenkins univariate dead band model to the case there are two responses of interest, possibly cross-correlated, and there are two controllable factors available to adjusting the process.
Multivariate extensions of the univariate dead band control models are evidently of practical interest, given that
most real-life processes have multiple responses to control and multiple controllable factors. The present paper is a first attempt in
 a particular case which is relatively tractable yet considerably useful in practice, when only two responses are influenced by two controllable factors. A recent paper by Govind et al. (2018) presents an approach for multivariate dead band control where the optimal threshold that balances the frequency of adjustments with the off-target cost is obtained from simulating the process for different threshold values. In the present paper, in contrast, we follow an analytical treatment of the problem that naturally generalizes the original Box-Jenkins derivations to the bivariate case. The paper is organized in several sections. Sections 2 and 3 present the assumptions behind the process and the assumptions behind the control policy (or ``controller"). Sections 4 and 5 discuss the loss function to be minimized and the cost assumptions involved. Section 6 gives the form of the optimal (dead band) bivariate policy. The optimal solution depends on knowing the second and fourth moments of a standardized bivariate time series, and these are derived in sections 7 and 8. With the moment formulae derived, an approximation to the loss function is given in section 9, and the accuracy of the approximation is studied in section 10. The numerical minimization of the loss function is addressed in section 11. This section contains a realistic scenario taken from the manufacturing of semiconductors where two responses are typically of interest.

%\section {The Process and Control Model.}
\section{The Process Model.}
By extension to the univariate Box-Jenkins machine tool model, pairs of disturbances $\; \z_t = (z_{ t,1}, z_{ t,2})' \; $
are assumed to follow a bivariate IMA(1,1) process, i.e.,
\be \label {FOR_PROCESS_10}
\z_ {t} - \z_{t-1} \;\;  =\;\;
\xalpha_{t} - {\bf \Theta} \xalpha_{ t-1} . \ee
Here, $\; {\bf \Theta} = (\theta_{ij})_{1\leq i,j \leq 2}\;$ is a known $2\times 2$ matrix,
and the pairs  $\;  \xalpha _t
= (\alpha_{t,1} , \alpha_{t, 2})'\; $ constitute a
bivariate cross correlated Gaussian white noise, i.e., normally distributed
pairs with stationary variance-covariance matrix
$$ {\bf C}_{\alpha} \quad = \quad \pmatrix{ \sigma_{1, \alpha }^2 & \kappa_ {\alpha}\cr
                                   \kappa_{\alpha} & \sigma_{2, \alpha}^2 }$$
for each
%$t\in \ZET$
time $t$ and without serial correlation, i.e.,
$\; \mbox {Cov} [\alpha_{t,l},  \alpha_{s,m}]=0\;$ for
$s\neq t$, $l,m \in \{1,2\}$.
The components of the variance-covariance matrix are assumed to be known.
For purposes of the control policy described in Section
\ref {The Control Model}, below, it is necessary to forecast the disturbances.
The minimum mean square error (MMSE) one step ahead forecast $\hat {\z}_ {t+1}$
computed at time $t$ for the disturbance vector $\z_{t+1}$ follows the EWMA recursion
\be \label {FOR_PROCESS_20}
\hat {\z}_ {t+1} \;\;  =\;\;
{\bf L} \z_{t,l} \,+\, ({\bf I} - {\bf L}) \hat {\z}_ {t} \ee
where $\; {\bf L} = {\bf I} - {\bf \Theta}$.
The vector of the one step ahead forecast errors for time $t+1$ is just
the white noise vector at time $t+1$, i.e.
\be \label {FOR_PROCESS_30}
{\z}_ {t+1} - \hat {\z}_ {t+1} \;\;  =\;\; \xalpha_{t+1}.\ee

\section{The Control Model.}
\label {The Control Model}
It is assumed that the process can be controlled via two {\em control factors}
$X_{s+1, 1}$, $X_{s+1, 2}$
which are set at the {\em adjustment  time} ({\em intervention time})
$s$, i.e., there is a delay of one time unit
until the adjustment takes effect.
$\X= (X_{s+1, 1}, X_{s+1, 2})'\; $ is the {\em control vector}.
The control variables
are supposed to compensate for the disturbances $z_ {t , 1}$,
$z_ {t  , 2}$ acting on the two process components at times $t=s+1, s+2,...$.
The vector of deviations from target under the effect of the
control variables is
$ \;
\mbox {\boldmath $d$}_t = \z_{t} -\X_{s+1}$, if no adjustments are made at  times $s+1,...,t$.
This model implies that a unit change on each control factor $X_{s+1,l}$ causes a unit change in the response, the deviations from target $d_{t,l}$.
In other words, the ``gain" matrix $\bf G$ in $ \;
\mbox {\boldmath $d$} _t = \z_{t} - \bf G \X_{s+1}$ equals the identity. There is no loss of generality with this, since if $\bf G$ is not the identity we simply use $\X_i={\bf G} \X_i^{(0)}$ in what follows, where $\X_i^{(0)}$ is the vector of original control factors.

\vspace{2mm}
An {\em intervention} into the process (or an adjustment) at time $s$
amounts to adjusting both control variables to values $X_{s+1, 1}$, $X_{s+1, 2}$.
In a manufacturing application, interventions provoke costs due to factors like
labor, material, process downtime and
loss of production volume. In the control of a disease though the supply of  a drug to a patient, the adjustment cost models the physical and emotional problems the patient may encounter that can be attributed to repeated applications of the drug. As will be shown in section 11, it is possible to determine a practical bivariate dead band policy without explicitly defining either the off-target or the adjustment costs (this has been emphasized by Box and Luce\~no, 1997, for the univariate case). Let $C>0$ be the cost of an intervention. This is a fixed adjustment cost regardless of the magnitude of the adjustment made. To reduce intervention costs it is reasonable not to intervene permanently
but only at selected intervention times $s$.

\vspace{2mm}
On the other hand, omitted adjustment leads to an increasing impact of
the disturbances, and consequently to increasing deviations from target and
increasing off target costs. In many cases the off target cost
can be measured as a linear function of the square deviation from target.
We assume costs $a_l>0$ per unit of the square deviation from target
in the $l$th production component, i.e., at times $t=s+1, s+2,...$ the off target
cost from component $l$ is
$\; a_l ( z_{t, l} -X_{s+1, l})^2 $.

\vspace {2mm}
In view of the off target cost it is reasonable to use the
predicted amount
of deviation from target as an intervention criterion.
Let the last adjustment be made at time $s$ with a resulting
adjustment vector
$\X_{s+1}$ at time $s+1$.
At times $s+k$, $k=1,2,...$ no intervention occurs as long as the vector of predicted deviations from target
%$\; \dx _{s+k+1} = ( \hat {d} _{s+k+1,1}, \hat {d} _{s+k+1,2})'\;$
%of the predicted deviations
\be \label {FOR_CONTROL_10}
\wh {\dx} _{s+k+1} \quad = \quad   \hat{\z}_{s+k+1} -\X_{s+1},\ee
is inside a noncritical region $D\subset \REL^2$ of the plane,
which we will refer to as a {\em dead area}.
%Over this time the control variables are constant at
At the first time $s+n$ with
\be \label {FOR_CONTROL_20}
 \hat {\dx} _{s+n+1 } \notin D
\; \;\; \mbox { or }\; \;\; n \geq n_0\ee
an alarm is given, and the control variables are adjusted so as to
compensate the predicted disturbance at time $s+n
+1$, i.e.,
$\; \X_{s+n+1} = \hat{\z}_{s+n+1}$.
The upper limit $n_0$  for the length of periods without adjustment
is prescribed for technical or security reasons, or it is a trivial upper limit,
e.g., the lifetime of machinery or production equipment.
In any case $n_0$ is a large upper limit, and alarms will generally result from the first
condition in formula (\ref {FOR_CONTROL_20}).
The random time of the next intervention after the last recorded
intervention time $s$ is
\be \label {FOR_CONTROL_30}
N\quad = \quad N_D\quad = \quad \min \Big\{ \min \{ n \in \NAT\, \vert \,
 \hat {\dx} _{s+n+1} \notin D \Big\}
, \; n_0\Big\} .\ee
Over the period $s+1, s+2, ...,s+N$ the control vectors remain constant at
$ \X_{s+1}$ and the vectors of the deviations from target are
\be \label {FOR_CONTROL_40}
{\z}_{s+1} -\X_{s+1},\; {\z}_{s+2} - \X_{s+1}, \;.\;.\;.,\;
{\z}_{s+N} - \X_{s+1}.\ee
The dead area $D\subset \REL^2$ in the bivariate case
corresponds to the univariate dead interval considered
 by Box and Jenkins (1963). Shifted along the time axis
the dead interval induces a {\em dead band}.
%Which is the appropriate shape of the dead area $D$ in the
%bivariate case? Recall that
%an alarm signal (\ref {FOR_CONTROL_20}) entails adjustments in
%{\em both} compensating variables. Hence neither of the
%two components $\hat {d} _{s+n+1,1}$, $ \hat {d} _{s+n+1,2}$ should
%have a higher inclination to provoke an alarm. Formally, this
%can be expressed by the following requirement:
%\begin {fett} {(D22)}
%\item [(DA)] At alarm time $N$, the bivariate distribution of the
%deviations from target
%$\hat {d} _{s+ N+1,1}$, $ \hat {d} _{s+N+1,2}$ for the
%two process components should be symmetric.
%\end {fett}
The appropriate shape of the dead area $D$ in the
bivariate case will be discussed in Section
\ref {The Standardized Loss Function}, below.

\section{The Loss Function.}
\label {The Loss Function}
A good control policy has to establish  a balance between the adjustment
cost and the off target cost.
Rare alarms reduce adjustment costs, but increase off target costs,
and vice versa.
From an economic point of view, the best policy is
the one which
minimizes the overall loss per time unit resulting from adjustments and
from being off target.
Under the assumptions of Section \ref {The Control Model},
a specific control policy is determined by the dead area $D \subset\REL^2$.
Hence we have to evaluate the loss incurred from running a process
under the policy described in
Section \ref {The Control Model} as a function $L(D)$ of the
dead area $D\subset \REL^2$.
%,  where $D$ is subject to therequirement (DA).

\vspace{2mm}
Consider a process run starting  at time 0, controlled
according to the policy described in Section \ref {The Control Model}.
Adjustments are made at the end of periods 1,2,3,... of random length
$N_1, N_2, N_3,...$ at times $S_1=N_1$, $S_2 = N_1 +N_2$,
$S_3 = N_1+N_2 +N_3$ and so on.
For each time unit $S_k+1,...,S_k+N_{k+1}$
in a period between two successive adjustments at times $S_k$ and
$S_{k+1} = S_k+N_{k+1}$
the off target cost is evaluated by the quadratic cost function
$\; a_l ( z_{S_k+i, l} -X_{S_k+1, l})^2 $.
Hence the overall loss per time unit in the $k$th period is
\be \label {FOR_LOSS_10}
V_k \quad = \quad \sum_{l=1}^2 a_l\sum_{ i =1}^ {N_{k+1} }
({z}_{S_k+i, l} -X_{S_k+1, l})^2 \;+\; C.\ee
For time $t$, let $K(t)$ be the number of periods
elapsed until time $t$. Then the loss per time unit until time $t$
is $\; V(t) = {1\over t} \sum _{k=1}^{K(t)} V_k$.
The process is assumed to run over a long time. Hence
it is reasonable to evaluate the expected overall
loss per time unit by the limit
$\; \lim _{t \to \infty} E[ V(t)]$.
To calculate the latter quantity we observe that
the pairs $(N_k, V_k)$, $k=1,2,...$ are serially independent
and identically distributed, i.e., they constitute a
{\em renewal reward process}, see the proof in appendix
\ref {The Renewal Reward Process Property}.
Hence an application of the {\em renewal reward theorem}, see  Ross
(1970), provides
the limit
$\; \lim _{t \to \infty} E[ V(t)]  =
{ E[V] \over E[N] }$. The expected hitting time $E[N]$ is what Box and Luce\~no (1997) call the {\em average adjustment interval}, or AAI. Calculating $E[V]$ from equation (\ref {FOR_LOSS_10}) we obtain the following loss function
\be \label {FOR_LOSS_20}
\begin {array} {l}
\displaystyle L(D) \quad = \quad
{ E[V] \over E[N] } \quad = \quad
{1\over E[N]} \sum_{l=1}^2 a_l\sum_{i=1}^N
E\Big[ ({z}_{i, l} -X_{1, l})^2 \Big] \;+\; {C \over E[N]} \quad = \quad
\\[2mm]
\displaystyle
{1\over E[N]} \sum_{l=1}^2 a_l\sum_{i=1}^N
E\Big[ ( \hat {z}_{i, l} -X_{1, l})^2 \Big] \;+\; {C \over E[N]}
\;+ \; a_1\sigma^2_{1, \alpha} \;+ \; a_2\sigma^2_ {2, \alpha}
\end {array}
\ee
as a function of the dead area $D \subset \REL^2$.
For the sake of convenience, in formula (\ref {FOR_LOSS_20})
and in subsequent calculations
we use the first period starting at time $1$
after adjustment at time $s=0$ to express the expectation $E[V]$.
$L$ depends on $D$ through the time $N=N_D$ between successive interventions, where
$N_D$ is defined by formula (\ref {FOR_CONTROL_30}).

\section{The Standardized Loss Function.}
\label {The Standardized Loss Function}
Using the loss function (\ref {FOR_LOSS_20}), we might define the
optimum control policy, i.e., the optimum dead area $D= D^{\star}$, as the
one which minimizes $L(D)$ over $D \subset \REL^2$.
%subject to the restriction (DA).
%However, up to now we have no restrictions on the shape of $D$.
However, determining an optimal solution
without restrictions on admissible shapes of the dead area
$D \subset \REL^2$ will be cumbersome.
By considering an appropriately standardized version of the loss
function $L(D)$ we get a more definite idea about
reasonable shapes of $D$. This will lead to a concise restriction
on $D$ which is appropriate for determining specific optimum
control policies.

\vspace {2mm}
The random variables $\hat {z}_{i, l} -X_{1, l}$, $l=1,2$,
which are necessary for calculating the loss function
$L(D)$ in formula (\ref {FOR_LOSS_20}), are
the components of the vectors $\; \hat {\z}_i - \X_1$.
From formulae (\ref {FOR_PROCESS_20}), (\ref {FOR_PROCESS_30}) and from
$\; \X_{1} = \hat {\z} _{1}\;$ we obtain
\ba \label {FOR_STANDARD_-20}
\hat {\z}_i - \X_1 \quad = \quad
\sum_{j=1}^{i-1} \bf L \xalpha_j.\ea
The components of the random vectors $\; \xbeta_j= {\bf L} \xalpha_j
= ( \beta_{j,1}, \beta_{j,2} ) ' \;$ have the
variance-covariance matrix
 \ba \label {FOR_STANDARD_-10}
\pmatrix{ \sigma_{1, \beta}^2 & \kappa_{\beta} \cr
           \kappa_{\beta} & \sigma_ {2, \beta}^2 }\quad = \quad
{\bf C} _{\beta} \quad = \quad \bf L {\bf C}_{\alpha} \bf L' \ea
and are serially uncorrelated. With respect to their cross-covariance ${\bf C}_{\beta}$, two cases have to be distinguished.

\vspace {2mm}
First, consider the case $\det {\bf C}_{\beta} = 0$. Then,
a linear relation holds between $\beta_{j,1}$ and
$\beta_{j,2}$ with probability 1, i.e., there exist
reals $c_1, c_2, c_3$ such that $\; \mbox{P} (
c_1 \beta_{j,1} + c_2 \beta_{j,2} = c_3 ) =1$,
see Schmetterer (2012). In this case, we are dealing essentially with a single univariate
problem which can be solved with the results of Box and Jenkins (1963).

\vspace {2mm}
In the sequel we assume $\det {\bf C} _{\beta} \neq  0$.
Then each random vector
$\; \xbeta_j = {\bf L} \xalpha_j
= ( \beta_{j,1}, \beta_{j,2} ) ' \;$ has a bivariate normal distribution
with variance-covariance matrix ${\bf C}_{\beta}$, see  Schmetterer
(2012).
The vectors  $ \ux_j = (u_{j,1} , u_{j, 2})'$
with $\, u_{ j,l} = {1\over \sigma_{l,\beta} } \beta_{j,l}\,$ constitute a
bivariate cross correlated (but serially uncorrelated) Gaussian unit white noise, i.e., normally distributed pairs with stationary variance-covariance matrix
$$ {\bf C}_u \quad = \quad \pmatrix{ 1 & \rho \cr
                                   \rho & 1 }$$
for each time $t$,
where $\; \rho = \rho_{\beta} = {\kappa_{\beta} \over \sigma_{1,\beta} \sigma_{2,\beta}}$,
and without serial correlation,
i.e., $\; \mbox {Cov} [u_{t,l},  u_{s,m}]=0\;$ for
$s\neq t$, $l,m \in \{1,2\}$. Note that assuming $\det {\bf C} _{\beta} \neq  0$ is equivalent to assuming $|\rho|<1$.

\begin{figure}
\begin{center}
\caption{Some possible forms of ``dead areas" on the plane.}
\label {GRA_DEAD_10}
\resizebox{18cm}{!} {\includegraphics{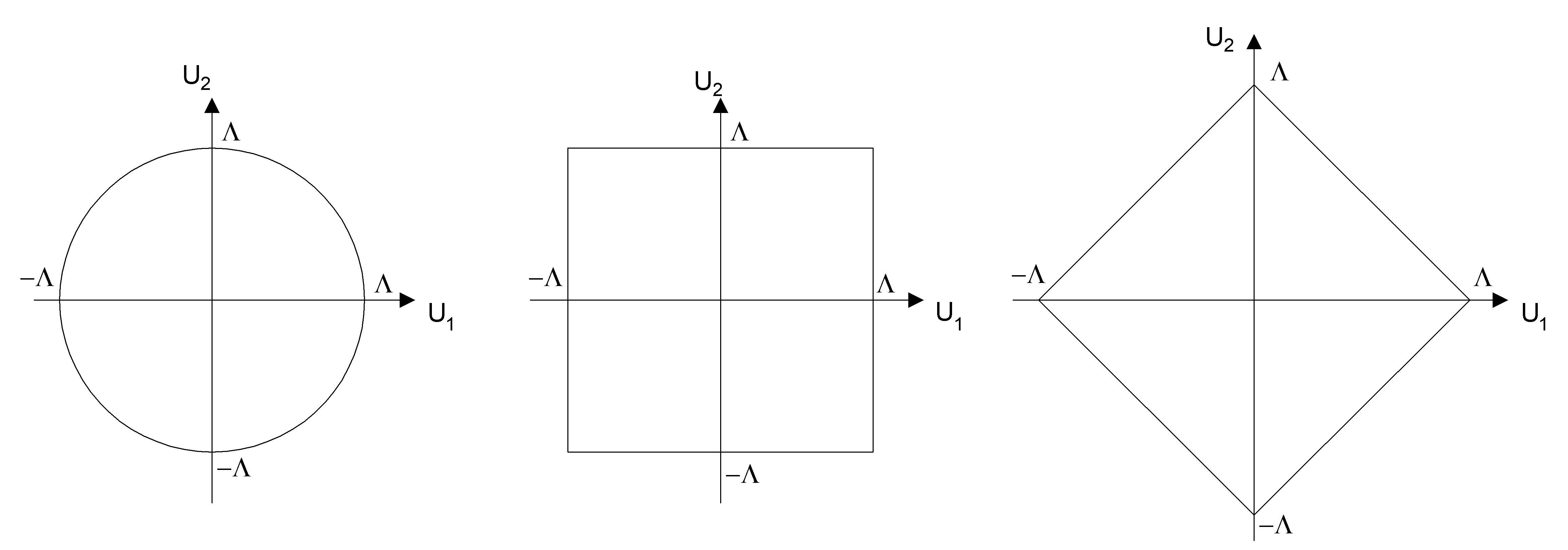}}
%\scalebox{1.5}{\includegraphics*[0,100][288,188]{gra_99_1.enc}}
%\scalebox{1.5}{\includegraphics{gra_0a0.enc}}
{\rule{6cm}{0.3mm}}
\end{center}
\end{figure}

 Letting $\; \UGR_i = \ux_1+...+\ux_i\;$ we obtain
from formula (\ref {FOR_STANDARD_-20})
\be \label {FOR_STANDARD_10}
  \hat {z} _{i,l} - X_{1,l}  \;\;=  \;\; \sigma_{l, \beta} U_{i-1,l} \quad
\mbox { for } \;i=1,2,..., \;\; l=1,2.
 \ee
Hence we can express the loss function $L(D)$ in the form
\be \label {FOR_STANDARD_20}
L(D) \quad = \quad
a_1\sigma^2_{1, \alpha} \;+ \; a_2\sigma^2_ {2, \alpha}
\;+\;
a_1 \sigma_{1,\beta}^2  G_1(D) \;+\;
a_2 \sigma_{2, \beta}^2 G_2(D) \;+\; {C\over E[N]},
\ee
where
\be \label {FOR_STANDARD_30}
G_l (D) \quad = \quad { E \Big [ \sum \limits _{j=1}^{N_D} U_{j-1,l}^2 \Big]
\over E[N_D] } \qquad
\mbox { for } \; l=1,2
\ee
will be referred to as the {\em scaled mean square deviation} (or MSD).
The predicted deviations from target
$\; \hat {d} _{k+1,l} =  \hat{z}_{k+1, l} -X_{1, l}\;$
(see equation \ref  {FOR_CONTROL_10}), can be expressed
as
$\; \hat {d} _{k+1,l} =   \sigma_{l, \beta}  U_{k,l}$. Hence,
equivalently to in (\ref  {FOR_CONTROL_20}),
the time $N$ of the next intervention is given by
\be \label {FOR_STANDARD_40}
N\quad = \quad N_{D'} \quad = \quad
\min \Big\{ \min \{ n \in \NAT\, \vert \, \UGR_n \notin D' \Big\},\; n_0\Big\}
 \ee
where
\be \label {FOR_STANDARD_50}
D' \quad = \quad \left \{
 \left( {x_1 \over \sigma_{1, \beta}}, {x_2 \over \sigma_{2,\beta}} \right)
\; \vert \; (x_1 , x_2) \in D \right\}.\ee
Now we are able to impose reasonable
restrictions on the shape of the dead area $D$.
Recall that
an alarm signal $\; \UGR_n \notin D'\;$  entails adjustments in
{\em both} compensating variables. Accordingly, neither of the
two components should
have a more prominent inclination to provoke an alarm.
Hence, since the bivariate distribution of the vectors
$\; \UGR_k = ( U_{ k,1}, U _{k,2})'\; $ is symmetric,
symmetricity should also hold for the dead area $D'$:
\begin {fett} {(D22)}
\item [(DA)]
 $D' \subset \REL^2$ should
be invariant under permutations $\; (x_1, x_2) \mapsto (x_2, x_1)\;$ of the
coordinates.
\end {fett}
To make the resulting control scheme practical for implementation in an industrial setting (the same could be argued for the control of some disease in a patient),
a further reasonable requirement is that $D'$ should be a convex area of a simple
geometric nature on the plane.
Three simple approaches to select $D'$ are shown in figure \ref {GRA_DEAD_10}:
a circle, a square, and a rotated square. Each of these areas conforms
to the above requirements, and each is a reasonable
adaptation of the univariate dead interval considered by
Box and Jenkins (1963) to the bivariate case.
From an economic point of view the best choice among these three
approaches is the one which guarantees a maximum dead area, i.e., a minimum of interventions,
at a prescribed level $\;c = L (D) = L(D')\;$ of the loss function.
We conjecture that in this sense the optimum shape is a circle.
However, determining the values $L(D')$ for circles $D'$ will be
difficult from a mathematical point  of view.
To provide a practical solution for application of the resulting control policy, we use a square shaped dead area $D'$.
In the following Section
\ref  {The Standardized Dead Area and the Optimum Control Policy} we
shall see
that  a rotated square $D'_{\Lambda}$ as on the
right-hand side of Figure \ref {GRA_DEAD_10} is most convenient for
calculations.

\section{The Standardized Dead Area and the Optimum Control Policy.}
\label {The Standardized Dead Area and the Optimum Control Policy}
\begin{figure}
\begin{center}
\caption{Dead rectangle with respect to $(W_{k,1}, W_{k,2})$ on the plane.}
\label {GRA_DEAD_20}
\resizebox{6cm}{!}{\includegraphics{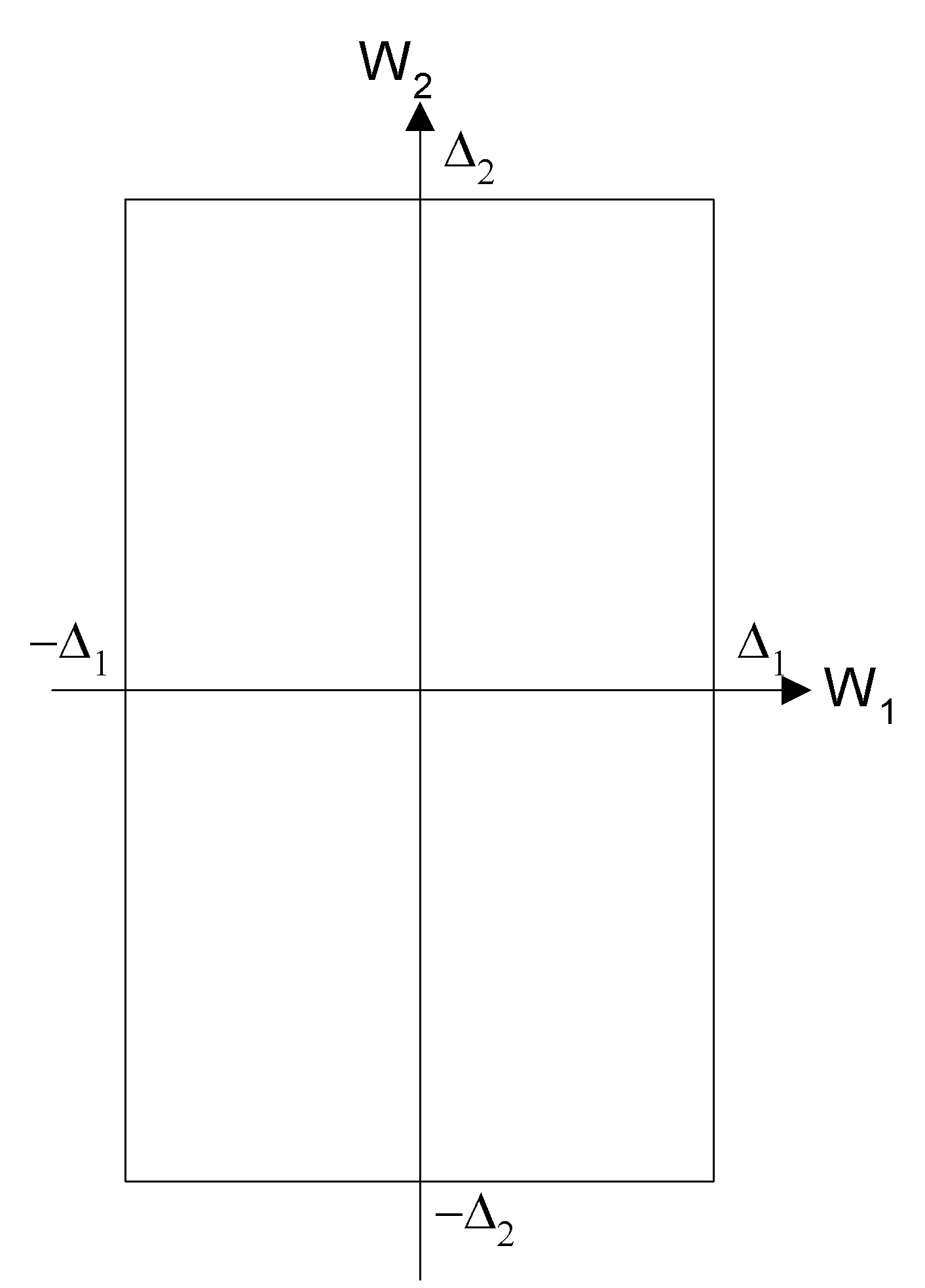}}
%\scalebox{1.5}{\includegraphics*[0,100][288,188]{gra_99_1.enc}}
%\scalebox{1.5}{\includegraphics{gra_0a0.enc}}

{\rule{6cm}{0.3mm}}
\end{center}
\end{figure}

As the dead area with respect to the standardized predicted deviations from
target
$( U _{k ,1}, U _{ k ,2})$, $k=1,2,...$ we
consider the interior
%$D'_{\Lambda}$
\be \label {FOR_STANDARD_60}
D'_{\Lambda} \quad = \quad \Big \{ (U_1, U_2) \; \vert \;
-\Lambda < U_1 + U_2 < \Lambda, \; -\Lambda < U_1 - U_2 < \Lambda
\Big\}
\ee
of a rotated square with vertices $(0, \Lambda)$,
$(\Lambda,0)$, $(0, -\Lambda)$, $(-\Lambda,0)$ as illustrated
by the right-hand side of Figure
\ref {GRA_DEAD_10}.
%Then
%\be \label {FOR_STANDARD_60}
%$$D'_{\Lambda} \quad = \quad \Big \{ (x_1, x_2) \; \vert \;
%-\Lambda < x_1 + x_2 < \Lambda, \; -\Lambda < x_1 - x_2 < \Lambda
%\Big\}.
%$$
%\ee
Hence the formula  (\ref {FOR_STANDARD_40})
for  the time $\; N= N_{\Lambda}= N_{D'_{\Lambda}}\; $ of the
next intervention amounts to
\be \label {FOR_STANDARD_70}
\begin {array}{l}
\displaystyle
N\quad = \quad N_{\Lambda} \quad = \quad \\[2mm]
\displaystyle \min \Big\{
\min \{ n \in \NAT\, \vert \,
 \vert W _{n ,1} \vert \geq {\Lambda \over \sqrt {2 (1+\rho) } } \;
\mbox { or } \;
 \; \vert W _{n ,2}  \vert \geq {\Lambda \over \sqrt {2 (1-\rho) }} \Big\} ,
\; n_0\Big\}
\end {array}  \ee
where
\be \label {FOR_STANDARD_80}
W_{k,1} \;\;=\;\; {U _{k ,1}+ U _{ k ,2} \over \sqrt {2 (1+\rho) } }
\;\;=\;\; w_{1,1}+...+ w_{k,1},  \quad w_{i,1} \;\; = \;\;
{ u_{i,1}+ u_{i,2} \over \sqrt {2 (1+\rho) } }, \ee
\be \label {FOR_STANDARD_90}
W_{k,2} \;\;=\;\; {U _{k ,1}- U _{ k ,2} \over \sqrt {2 (1-\rho) } }
\;\;=\;\; w_{1,2}+...+ w_{k,w},  \quad w_{i, 2} \;\; = \;\;
{ u_{i,1}- u_{i,2} \over \sqrt {2 (1-\rho) } }, \ee
%\be \label {FOR_STANDARD_100}
%\Delta_1 \;\;=\;\; {\Lambda \over \sqrt {2 (\rho+1) } }, \quad
%\Delta_2 \;\;=\;\; {\Lambda \over \sqrt {2 (\rho-1) } }. \ee
It is easy to verify that the
pairs $ (w_{t,1} , w_{t, 2})$
constitute a bivariate
uncorrelated Gaussian unit white noise, i.e., normally distributed
pairs with stationary variance-covariance matrix
$$ {\bf C}_w \quad = \quad \pmatrix{ 1 & 0 \cr
                                    0& 1 }$$
for each time $t$ and without serial correlation, i.e.,
$\; \mbox {Cov} [w_{t,l},  w_{s,l}]=0\;$ for
$s\neq t$, $l=1,2$,
$\; \mbox {Cov} [w_{t,1},  w_{s,2}]=0\;$ for
$s\neq t$.
By formula (\ref {FOR_STANDARD_70}), the time $N= N_{\Lambda} $
of the next intervention is expressed as the first exit time
of the cross-independent bivariate random walk
$(W_{k,1}, W_{k,2})$  from the  open
rectangle $\, (-\Delta_1; \Delta_1) \times
(- \Delta_2; \Delta_2)$,
where  $\; \Delta_1 = {\Lambda \over \sqrt {2 (1+ \rho) } }$,
$\; \Delta_2 = {\Lambda \over \sqrt {2 (1-\rho) } }$, as
illustrated by  figure \ref {GRA_DEAD_20}.

\vspace {2mm}
Hence we have two equivalent descriptions of the dead area:
\begin {itemize}
\item The dead area with respect to the standardized observations
$( U _{k ,1}, U _{ k ,2})$, $k=1,2,...$, is  the interior
$D'_{\Lambda}$  of a rotated rectangle as defined by formula
(\ref {FOR_STANDARD_60}).
\item
The dead area with respect to the uncorrelated pairs
$( W _{k ,1}, W _{ k ,2})$, $k=1,2,...$, defined by
formulae (\ref {FOR_STANDARD_80}) and (\ref {FOR_STANDARD_90})
is  the interior
\be \label {FOR_STANDARD_110}
D''_{\Lambda} \quad = \quad
\quad \Big \{ (y_1, y_2) \; \vert \;
\vert y_1 \vert < {\Lambda \over \sqrt {2 (1+\rho) } }, \; \vert y_2 \vert
< {\Lambda \over \sqrt {2 (1-\rho) } }
\Big\}
\ee
of a rectangle parallel to the axes.
 \end {itemize}
%The two definitions of the dead area are equivalent
%if the parameters $\Lambda$ and  $\Delta_1$, $\Delta_2$ are related
% by  formula (\ref {FOR_STANDARD_100}).
%For these values
The loss function can be expressed as a function
$$ L(\Lambda) \;\;=\;\; L( D'_{\Lambda}) \;\;=\;\; L( D''_{\Lambda})
$$
of the parameter $\Lambda$ ranging over $(0;+\infty)$.
Hence the optimum
control policy can be defined by a value $\Lambda^{\star}$ which
minimizes $L(\Lambda)$ for all $\Lambda >0$.
%By Calculations on $L(\Lambda)$ are simplified by the
%By symmetry, $G_1 (D_\lambda
In analogy to the univariate case, the dead areas $D ' _{\Lambda}$ (rotated square centered
in the origin) and $D''_{\Lambda}$ (rectangle parallel to the axes,
centered in the origin), when shifted along the time axis induce {\em dead bars}. From a practical point of view, displaying $U_{t,1}+U_{t,2}$ and $U_{t,1}-U_{t,2}$ on ``adjustment" charts with limits at $\pm \Lambda$ is probably preferred, as we illustrate in Section 11. We first consider the moments needed to compute the standardized loss function.

\section{Relations among Moments of {\boldmath $U_{N,l}$ and $W_{N,m}$}.}
\label {Relations among Moments}
To evaluate the loss function $\, L(D'_{\Lambda}) =L( \Lambda)\,$ we need the moments
$E[U_{N,l}^2]$, $E[U_{N,l}^4]$ of the
standardized accumulated deviations
from target, which we use for this purpose in Section
\ref {Asymptotic Evaluation of the Loss Function}.
By the choice of the dead area $D_{\Lambda}$, $U_{N,1}$ and $U_{N,2}$ have the
same distribution. Hence
\be \label {FOR_RELATIONS_10}
E[U_{N,1}^q] \;\;=\;\; E[U_{N,2}^q] \quad \mbox{ for } \; q\in \NATO. \ee
From the symmetry of the underlying bivariate normal distribution
and from the symmetry of the dead area it is clear that
\be \label {FOR_RELATIONS_20}
E[U_{N,1}^q] \;\;=\;\; 0\;\;=\;\;  E[U_{N,2}^q] \quad \mbox{ for } \; q =1,3,5,.... \ee
Because of the correlation among the variables
$U_{k,1}$, $U_{k,2}$, direct calculation of the moments
$\ E[U_{N,l}^q]$, $ q= 2,4,...$, is rather involved. It is more convenient to calculate
the moments $E[W_{N,m}^q]$, see Section \ref {Moments of}, and then
to derive $E[U_{N,l}^2]$ and $E[U_{N,l}^4]$.
For this purpose, we establish relations
among the moments of $U_{N,l}$ and the moments of $W_{N,m}$.
From formulae (\ref {FOR_STANDARD_80}) and (\ref {FOR_STANDARD_90}) we obtain
\be \label {FOR_RELATIONS_30}
E[U_{N,l}^2] \quad = \quad
{1\over 2} \Big \{ (1+\rho) E[W_{N,1}^2] \;+ \; (1-\rho) E[W_{N,2}^2] \Big\}, \ee
\be \label {FOR_RELATIONS_40}
E[U_{N,l}^4] \;+\; 3 E[ U_{N,1}^2 U_{N,2}^2 ] \quad = \quad
 (1+\rho) ^2 E[W_{N,1}^4] \;+ \; (1-\rho)^2 E[W_{N,2}^4] , \ee
\be \label {FOR_RELATIONS_50}
E[U_{N,l}^4] \;-\;  E[ U_{N,1}^2 U_{N,2}^2 ] \quad = \quad
 2(1-\rho ^2)  E[W_{N,1}^2 W_{N,2}^2]. \ee
Combining equations (\ref {FOR_RELATIONS_40}) and (\ref {FOR_RELATIONS_50})
we obtain
\be \label {FOR_RELATIONS_60}
E[U_{N,l}^4]  \quad = \quad
{ 1\over 4} \Big \{ (1+\rho)^2 E[W_{N,1}^4] \;+\;
6(1-\rho ^2)  E[W_{N,1}^2 W_{N,2}^2]\; +\;
 \; (1-\rho)^2 E[W_{N,2}^4 ] \Big\}. \ee

\section{Moments of \boldmath{$W_{N,m}$}.}
\label {Moments of}
Because of the independence of the variables $ W_{k,1}$, $W_{k,2}$
we can adapt a method used by  Box and Jenkins (1963) for the univariate
case to determine an approximation of the moments
$ E[ W_{N,m}^q]$ and of $ E[ W_{N,1}^p W_{N,2}^r]$.
In this derivation, we ignore the
upper limit $n_0$  for the length of periods without adjustment.
See the explanation on
$n_0$ in Section \ref {The Control Model}.

\vspace {2mm}
From the symmetry of the dead area and of the underlying normal
distribution it is clear that
\be \label {FOR_MOMENTSOF_10}
E[W_{N,1}^q] \;\;=\;\; 0\;\;=\;\;  E[W_{N,2}^q] \quad \mbox{ for } \; q =1,3,5,.... \ee
As in Section \ref {The Standardized Dead Area and the Optimum Control Policy}
we use the abbreviating notation
$\; \Delta_1 = {\Lambda \over \sqrt {2 (1+\rho) } }$,
$\; \Delta_2 = {\Lambda \over \sqrt {2 (1-\rho) } }$.
For $k \in \NAT$, let $h_k \colon \REL \to \REL$ be the joint density of
$ W_{1,m}, ..., W_{k,m}$, and let $g_{k,m} \colon \REL \to \REL$ be defined
by
\be \label {FOR_MOMENTSOF_20}
g_{k,m} (y) \quad = \quad \int \limits_{ \vert y_1 \vert < \Delta_m,
...,  \vert y_{k-1} \vert < \Delta_m}
h_k ( y_1, ...,y_{k-1}, y)\, \mbox{d}y_1... \mbox{d}y_{k-1}.
\ee
Since $\; W_{k,m} = w_{k,m}+ W_{k-1,m}\;$  with $w_{k,m}$ distributed according
to $N(0,1)$, the functions $g_{k,m}$ follow the recursion
\be \label {FOR_MOMENTSOF_30}
g_{k,m} (x_m) \quad = \quad \int _{ \vert y_m \vert < \Delta_m}
g_{k-1,1} (y_m) \varphi( x_m -y_m) \, \mbox{d}y_m,
\ee
where
$\; \varphi( z) = {1\over \sqrt {2\pi}} \exp \left (
{- z^2 \over 2} \right)\;$ is the density function of the
normal distribution $N(0,1)$.  $g_{k,m} $ is a  density of
$W_{k,m}$ in the event $\{ \vert W_{1,m}\vert < \Delta_m,...,
\vert W_{k-1,m}\vert < \Delta_m\} $, i.e.,
\be \label {FOR_MOMENTSOF_40}
%\mbox{P} (W_{n,m} \in B, N\geq n) \quad = \quad
\mbox{P} (W_{ k,m} \in B, \; \vert W_{1,m}\vert < \Delta_m,...,
\vert W_{ k-1,m}\vert < \Delta_m ) \quad = \quad
\int_ B g_{ k,m} (x_m) \, \mbox{d} x_m \ee
for Borel sets $B$.
From formula (\ref {FOR_MOMENTSOF_40}) it follows that
$$ { g_{n-1,1} \cdot g_{ n-1, 2} \cdot \EINS_{ ( -\Delta_1; \Delta_1)
\times ( -\Delta_2; \Delta_2) } \over \mbox{P} ( N\geq n) } $$
is a joint conditional density of $W_{ n-1 ,1}$ and
$W_{ n-1, 2}$ under the condition $N\geq n$, where
$\EINS_B$ is the indicator function of a set $B$.
We follow the intuitively reasonable approach used by
Box and Jenkins (1963) for the univariate case: We approximate
the joint conditional distribution of
$W_{ n-1 ,1}$ and $W_{ n-1, 2}$ under the condition $N\geq n$ by a
bivariate uniform distribution over the dead rectangle
$ \,  ( -\Delta_1; \Delta_1)
\times ( -\Delta_2; \Delta_2) $, i.e., we assume
\be \label {FOR_MOMENTSOF_50}
{ g_{n-1,1} \cdot g_{ n-1, 2} \cdot \EINS_{ ( -\Delta_1; \Delta_1)
\times ( -\Delta_2; \Delta_2) } \over \mbox{P} ( N\geq n) }
\quad \approx \quad { 1\over 4 \; \Delta_1 \; \Delta_2}.
\ee
The accuracy of this approximation is studied in Section 10.

Under this approximation we obtain for $p,r\in \NATO$
$$ E[ W_{N,1}^ p W_{N,2}^ r \vert N=n ] \; \mbox{P} (N=n) \quad =
_{ (\ref {FOR_STANDARD_70}), \; (\ref {FOR_MOMENTSOF_40}) }  \quad $$
$$
%\int \limits _ { \vert x_1 \vert \geq \Delta_1}
%\int \limits _ {  x_2 \in \REL}
\int \limits _ { \begin {array} {l}
 \scriptstyle \{ \vert x_1 \vert \geq \Delta_1 \} \cup \\
\scriptstyle
\{
%\vert x_1 \vert < \Delta_1,
\vert x_2 \vert \geq \Delta_2  \}
\end {array} }
x_1^p x_2 ^r g_{ n,1 } (x_1)
g_{ n, 2 } (x_2) \, \mbox{d} x_1 \mbox{d} x_2
\quad =
_{ (\ref {FOR_MOMENTSOF_30}) }  \quad
$$
$$
%\int\limits _ { \vert x_1 \vert \geq \Delta_1}
%\int\limits _ {  x_2 \in \REL}
%\int \limits _ { \{ \vert x_1 \vert \geq \Delta_1 \} \cup
%\{ \vert x_1 \vert < \Delta_1, \vert x_2 \vert \geq \Delta_2  \}}
\int \limits _ { \begin {array} {l}
 \scriptstyle \{ \vert x_1 \vert \geq \Delta_1 \} \cup \\
\scriptstyle
\{
%\vert x_1 \vert < \Delta_1,
\vert x_2 \vert \geq \Delta_2  \}
\end {array} }
x_1^p x_2 ^r
\int\limits  _{ \vert y_1 \vert < \Delta_1}
\int\limits  _{ \vert y_2 \vert < \Delta_2}
g_{n-1,1} (y_1) g_{n-1,2} (y_2)  \varphi( x_1 -y_1) \varphi( x_2 -y_2 )
\, \mbox{d}y_2  \mbox{d}y_1 \, \mbox{d} x_1 \mbox{d} x_2 \quad
\approx _{ (\ref {FOR_MOMENTSOF_50}) }  \quad
$$
$$
{ \mbox {P} (N \geq n) \over 4 \; \Delta_1 \; \Delta_2} \;\;
\Bigg  \{
\int \limits _ { \vert x_1 \vert \geq \Delta_1} x_1 ^p
\int _ {-\Delta_1 - x_1 } ^{ \Delta_1 -x_1} \varphi(y_1)
\mbox{d}y_1 \, \mbox{d} x_1  \; \cdot
\int \limits _ {  x_2 \in \REL} x_2 ^r
\int _ {-\Delta_2 - x_2 } ^{ \Delta_2 -x_2} \varphi(y_2)
\mbox{d}y_2 \, \mbox{d} x_2 \;\;\; \; + \;\;
$$
$$ \leftline{ \hfill $\displaystyle
\int \limits _ { \vert x_1 \vert < \Delta_1} x_1 ^p
\int _ {-\Delta_1 - x_1 } ^{ \Delta_1 -x_1} \varphi(y_1)
\mbox{d}y_1 \, \mbox{d} x_1  \; \cdot
\int \limits _ {  \vert x_2 \vert \geq \Delta_2 } x_2 ^r
\int _ {-\Delta_2 - x_2 } ^{ \Delta_2 -x_2} \varphi(y_2)
\mbox{d}y_2 \, \mbox{d} x_2 \Bigg\} \quad =
$ \hspace*{4 mm} } $$
$$ { \mbox {P} (N \geq n) \over 4 \; \Delta_1 \; \Delta_2} \;
\bigg  \{ I(\Delta_1, p) \Big [ I(\Delta_2, r) + J(\Delta_2, r)\Big] \;+\;
J(\Delta_1, p)  I(\Delta_2, r) \bigg\},$$
where for $\Delta\geq 0$, $ q\in \NATO$
\be \label {FOR_MOMENTSOF_I_INTEGRAL}
I (\Delta, q) \;\; = \;\; \int _ { \vert x\vert \geq \Delta}
x^q \Big[ \phi (x+\Delta) - \phi (x-\Delta)\Big] \, \mbox{d} x, \ee
\be \label {FOR_MOMENTSOF_J_INTEGRAL}
J (\Delta, q) \;\; = \;\; \int _ { \vert x\vert < \Delta}
x^q \Big[ \phi (x+\Delta) - \phi (x-\Delta)\Big] \, \mbox{d} x. \ee
The integrals $I (\Delta, q)$ and $ J (\Delta, q)$ are evaluated by
proposition
\ref {Integrals of the Normal Distribution Function}.\ref {PROPO_APPENDIX2_10} in the
appendix \ref {Integrals of the Normal Distribution Function}.
In particular,
\be \label {FOR_MOMENTSOF_60}
I (\Delta, 0) \;\; = \;\; 2 \Big\{
\varphi (0) - \varphi( 2\Delta) + 2 \Delta \Big[1- \phi (2\Delta)\Big] \Big\}, \quad
J(\Delta, 0) \;\;=\;\; 2 \Delta - I(\Delta, 0).\ee
Letting $p= 0=r$ into the above derivation we obtain from
(\ref {FOR_MOMENTSOF_60})
$$ \mbox{P} (N=n) \quad \approx \quad
{ \mbox {P} (N \geq n) \over 4 \; \Delta_1 \; \Delta_2} \;
\bigg  \{ 2\Delta_2 I(\Delta_1, 0)  \;+\;
\Big[ 2\Delta_ 1 - I(\Delta_1, 0) \Big]  I(\Delta_2, 0) \bigg\}.$$
Finally for $p, r \in \NATO$
\be \label {FOR_MOMENTSOF_70}
\begin {array} {l}
\displaystyle E[ W_{N,1}^ p W_{N,2}^ r ] \quad =
  \quad \sum _{n=1}^ {\infty}
E[ W_{N,1}^ p W_{N,2}^ r \vert N=n ] \; \mbox{P} (N=n) \quad \approx
  \quad \\[3mm]
\displaystyle
{ I(\Delta_1, p) \Big [ I(\Delta_2, r) + J(\Delta_2, r)\Big] \;+\;
J(\Delta_1, p)  I(\Delta_2, r) \over
2\Delta_2 I(\Delta_1, 0)  \;+\;
\Big[ 2\Delta_ 1 - I(\Delta_1, 0) \Big]  I(\Delta_2, 0)}\;\;.
\end {array}
\ee
In particular for $q \in \NATO$
\be \label {FOR_MOMENTSOF_80}
E[ W_{N,1}^ q ] \quad \approx
  \quad
{ I(\Delta_1, q) 2\Delta _2  \;+\;
J(\Delta_1, q)  I(\Delta_2, 0) \over
2\Delta_2 I(\Delta_1, 0)  \;+\;
\Big[ 2\Delta_ 1 - I(\Delta_1, 0) \Big]  I(\Delta_2, 0)}\;\;.
\ee

\section{An Approximation of the Loss Function
{\boldmath $L(\Lambda)$}.}
\label {Asymptotic Evaluation of the Loss Function}
%Analogousl
The standardized accumulated deviations from target
$\, \sum  _{j=1}^{N_D} U_{j-1,1}^2 \,$ and
$\, \sum  _{j=1}^{N_D} U_{j-1,2}^2 \,$ have the same distribution.
Hence from formula (\ref {FOR_STANDARD_30})
$\; G_1 ( D_{\Lambda}' ) =  G_2 ( D_{\Lambda}')$.
Consider
the martingales
$(R_{n,l}) _{n\in \NAT} $, $(Y_{n,l}) _{n\in \NAT}$, $(Z_{n,l}) _{n\in \NAT}$
defined by formulae (\ref {FOR_MARTINGALE_10}) and
(\ref {FOR_MARTINGALE_20}) in Appendix
\ref {Some Useful Martingales}.
Obviously, $N_D$ is a stopping time for these martingales,
uniformly bounded by $N_D \leq n_0$.
Hence the optional stopping theorem, see
Rogers and Williams (1994), provides
\be \label {FOR_LOSSAPPROXIMATION_10}
E[N] \;\; =  \;\; E[ U_{N,l}^2], \qquad
E\Big [ \sum \limits _{j=1}^{N_D} U_{j-1, l}^2 \Big] \;\; = \;\;
{ E[ U_{N,l}^4] \over 6} \,-\, {E[N] \over 2}. \ee
We point out that these expressions are exact and not approximations, as suggested by Box and Jenkins (1963).

Inserting into formula  (\ref {FOR_STANDARD_30}) we obtain the scaled MSD:
\be \label {FOR_LOSSAPPROXIMATION_20}
G_1 ( D_{\Lambda}' ) \quad =  \quad G_2 ( D_{\Lambda}')\quad = \quad
{ E[ U_{N,l}^4] \over 6  E[ U_{N,l}^2] } \;-\; {1\over 2}. \ee
Hence, from formula  (\ref {FOR_STANDARD_20}) we have that
\be \label {FOR_LOSSAPPROXIMATION_30}
\begin {array} {l}
\displaystyle L(\Lambda) \quad = \quad L( D_{\Lambda}' ) \quad = \quad \\[3mm]
\displaystyle
a_1 \sigma_{1, \alpha}^2 \;+\; a_2 \sigma_{2, \alpha}^2 \;+\;
( a_1 \sigma_{1,\beta}^2 + a_2 \sigma_{2, \beta} ^2  )
\left ( { E[ U_{N,l}^4] \over 6  E[ U_{N,l}^2] } \;-\; {1\over 2}\right)\;
+\; {C\over E[ U_{N,l}^2]}
\;. \end {array}
\ee
To obtain an approximation of the loss function $\; L(\Lambda)  =  L( D_{\Lambda}' )$,
we insert into formula (\ref {FOR_LOSSAPPROXIMATION_30})
the approximations for the moments
$E[ U_{N,l}^2]$ and $ E[ U_{N,l}^4]$  determined from
formulae (\ref {FOR_RELATIONS_30}), (\ref {FOR_RELATIONS_60}),
(\ref {FOR_MOMENTSOF_70}), (\ref {FOR_MOMENTSOF_80}).

\section{Accuracy of the approximations}

The expressions for the moments (\ref {FOR_RELATIONS_30}) and (\ref {FOR_RELATIONS_60}) are based on the approximation (\ref{FOR_MOMENTSOF_50}). Similarly as what Box and Jenkins (1963) reported for the univariate case, the assumption of a uniform distribution for the standardized bivariate process
$\mbox{\boldmath $W$} _{n-1}$ before the process falls out of the dead area was found to be inaccurate, particularly for large values of $\Lambda$. The geometrical reason for this problem is that, for large $\Lambda$, the points $(W_{n-1,1}, W_{n-1,2})$ will gather closer to the boundaries of the dead area than to the center of the region. Therefore, a correction regression equation was developed empirically by computing, through simulation, the ``real" moments $E_r[U_{N,l}^2]$ and $E_r[U_{N,l}^4]$ and computing the differences $D_2=E[U_{N,l}^2]-E_r[U_{N,l}^2]$ and $D_4=E[U_{N,l}^4]-E_r[U_{N,l}^4]$. Here, $E_r[U_{N,l}^2]$ and $E_r[U_{N,l}^4]$ were estimated by simulating 50,000 renewals for $|\rho| \in \{0.1,0.25,0.5,0.75,0.85,0.95\}$ and $\Lambda \in \{1,2,...,15\}$. The moments $E[U_{N,l}^2]$ and $E[U_{N,l}^4]$ were computed as in (\ref {FOR_RELATIONS_30}) and (\ref {FOR_RELATIONS_60}).
Note from
%(\ref {FOR_RELATIONS_30}-\ref {FOR_RELATIONS_60}) and
(\ref{FOR_STANDARD_70}-\ref{FOR_STANDARD_90}) that the moments are invariant with respect to the sign of the cross-correlation coefficient $\rho$.

The following correction model was fitted to the errors in the second order moment data:
\be
\label{D2}
D_2^{0.337} = 0.385 + 0.133 \Lambda -0.840 |\rho| - 0.00172 \Lambda^2 + 0.90 \rho^2 + 0.0375 \Lambda |\rho|
\ee
that is, a quadratic polynomial model in $\Lambda$ and $\rho$ was fitted after a Box-Cox power transformation was applied to the data (hence the exponent in the left hand side). This model was fitted for $\Lambda>2$ since for small values of $\Lambda$ the analytic formula  provides a good approximation to the real moment. Fortunately, model (\ref{D2}) provides an excellent fit, with $R^2=0.997$ and the p-values associated with the tests for the significance of each regressor equal to zero up to three decimal places in all cases.

For the errors in the fourth order moments, the corresponding fitted model was:
\be
\label{D4}
D_4^{0.224} = 0.515+0.386 \Lambda -1.25 |\rho| + 0.00118 \Lambda^2 + 1.44 \rho^2 + 0.0768 \Lambda |\rho|
\ee
where similarly as before, a full quadratic polynomial in $\Lambda$ and $\rho$ was fitted after a Box-Cox transformation was applied to the errors. Values $\Lambda\leq 2$ were excluded from the regression, similarly as before. The fit again is excellent, giving $R^2=0.999$ and all p-values of the individual tests of significance for each model parameter smaller or equal to 0.001.

In order to minimize the standardized cost function, the correction formulae $(\ref{D2}-\ref{D4})$ were used for $\Lambda \geq 2$. For $\Lambda<2$, no correction was used and the analytical formulae (\ref {FOR_RELATIONS_30}), (\ref {FOR_RELATIONS_60}) were directly utilized instead.

\section{Minimization of the standardized loss function}

From (\ref{FOR_LOSSAPPROXIMATION_30}), it is evident that the optimal solution $\Lambda^*$ depends on the relative cost parameter
\[ C'=\frac{C}{a_1 \sigma_{1,\beta}^2 + a_2 \sigma_{2,\beta}^2}. \] The only other parameter that the optimal solution depends on is the value of $|\rho|$, the cross-correlation of the bivariate series $\xbeta_j = \bf L \xalpha_j$. To find $\Lambda^*$, the cost function
\[
L'(\Lambda)=\frac{L(\Lambda)}{a_1 \sigma_{1,\beta}^2 + a_2 \sigma_{2,\beta}^2}-\frac{a_1 \sigma_{1,\alpha}^2+a_2 \sigma_{2,\alpha}^2}{a_1 \sigma_{1,\beta}^2 + a_2 \sigma_{2,\beta}^2}
= \frac{E[U^4_{N,l}]}{6 E[U^2_{N,l}]} - \frac{1}{2}+\frac{C'}{E[U^2_{N,l}]}
\]
 was minimized using Matlab's \texttt{fminbnd} function, which minimizes a non-linear function subject to bounds (bounds of $0.1$ and $20$ were used in all cases in the table below). For $\Lambda>2$, the two moments were corrected using (\ref{D2}-\ref{D4}). The solutions reported in this section were confirmed to provide the unique minimizer of the function within the interval (the Matlab code used in this section is available from the first author upon request).

Table 1 shows the optimal solution $\Lambda^*$, the corresponding value of the loss function $L'(\Lambda^*)$, the scaled MSD value $(G_l(D'_{\Lambda^*}))$, and the Average Adjustment Interval (AAI$=E[N]$) for a variety of values of $|\rho|$ and $C'$. From it, a potential user can select a solution by finding acceptable MSD and AAI values, without having to define an explicit cost $C'$.  In general terms, the optimal limit $\Lambda^*$ increases with increasing relative fixed adjustment cost ($C'$) and with increasing correlation ($|\rho|$). The cost function was observed to be fairly flat around the minimum point, so small departures of $\Lambda$ from the optimum value $\Lambda^*$ will not be of practical importance.\\

\begin{center}
\begin{scriptsize}
\begin{tabular}{rrrrrr}
\hline
\multicolumn{6}{c}{Table 1. Some optimal solutions.}\\
\hline
{$C'$}&{$|\rho|$}&{$\Lambda^*$}&{$L'(\Lambda^*)$}&{Scaled MSD}&{AAI}\\
\hline
1  &  0.0  &  1.01  &  0.63  &  0.10  &  1.91  \\
1  &  0.3  &  1.02  &  0.63  &  0.11  &  1.91  \\
1  &  0.6  &  1.21  &  0.63  &  0.16  &  2.11  \\
1  &  0.9  &  1.39  &  0.54  &  0.22  &  2.69  \\
4  &  0.0  &  2.85  &  1.66  &  0.69  &  4.10  \\
4  &  0.3  &  2.73  &  1.69  &  0.71  &  4.06  \\
4  &  0.6  &  2.73  &  1.65  &  0.68  &  4.12  \\
4  &  0.9  &  2.99  &  1.44  &  0.56  &  4.57  \\
7  &  0.0  &  3.40  &  2.30  &  0.99  &  5.35  \\
7  &  0.3  &  3.29  &  2.33  &  1.02  &  5.33  \\
7  &  0.6  &  3.33  &  2.28  &  1.00  &  5.47  \\
7  &  0.9  &  3.61  &  2.00  &  0.84  &  6.01  \\
10  &  0.0  &  3.79  &  2.81  &  1.24  &  6.35  \\
10  &  0.3  &  3.69  &  2.85  &  1.27  &  6.35  \\
10  &  0.6  &  3.74  &  2.78  &  1.24  &  6.51  \\
10  &  0.9  &  4.04  &  2.46  &  1.06  &  7.14  \\
20  &  0.0  &  4.66  &  4.12  &  1.87  &  8.88  \\
20  &  0.3  &  4.58  &  4.15  &  1.92  &  8.93  \\
20  &  0.6  &  4.64  &  4.05  &  1.87  &  9.17  \\
20  &  0.9  &  5.00  &  3.63  &  1.63  &  10.02  \\
50  &  0.0  &  6.06  &  6.76  &  3.16  &  13.91  \\
50  &  0.3  &  6.00  &  6.76  &  3.21  &  14.06  \\
50  &  0.6  &  6.11  &  6.59  &  3.14  &  14.47  \\
50  &  0.9  &  6.56  &  5.95  &  2.78  &  15.75  \\
80  &  0.0  &  6.91  &  8.66  &  4.10  &  17.54  \\
80  &  0.3  &  6.86  &  8.65  &  4.14  &  17.76  \\
80  &  0.6  &  7.01  &  8.42  &  4.05  &  18.30  \\
80  &  0.9  &  7.52  &  7.64  &  3.62  &  19.90  \\
100  &  0.0  &  7.35  &  9.74  &  4.63  &  19.58  \\
100  &  0.3  &  7.31  &  9.71  &  4.67  &  19.85  \\
100  &  0.6  &  7.47  &  9.45  &  4.57  &  20.47  \\
100  &  0.9  &  8.02  &  8.59  &  4.09  &  22.24  \\
400  &  0.0  &  10.71  &  19.98  &  9.73  &  39.04  \\
400  &  0.3  &  10.74  &  19.79  &  9.72  &  39.74  \\
400  &  0.6  &  11.04  &  19.21  &  9.48  &  41.13  \\
400  &  0.9  &  11.89  &  17.56  &  8.62  &  44.75  \\
700  &  0.0  &  12.44  &  26.59  &  13.06  &  51.71  \\
700  &  0.3  &  12.52  &  26.28  &  13.00  &  52.70  \\
700  &  0.6  &  12.89  &  25.47  &  12.66  &  54.64  \\
700  &  0.9  &  13.90  &  23.31  &  11.56  &  59.57  \\
1000  &  0.0  &  13.67  &  31.87  &  15.72  &  61.91  \\
1000  &  0.3  &  13.79  &  31.46  &  15.62  &  63.15  \\
1000  &  0.6  &  14.22  &  30.47  &  15.21  &  65.55  \\
1000  &  0.9  &  15.36  &  27.89  &  13.92  &  71.58  \\
\hline
\end{tabular}
\end{scriptsize}
\end{center}
\newpage

\noindent{\bf Relation with Box and Jenkins' univariate optimal solution}\\

Clearly, our formulation reduces to solving two separate univariate problems using Box and Jenkins (1963) formulation, one for each response and each controllable factor $l$, when $\kappa_{\alpha}=0$ (which implies $\rho=0$) and both $\bf \Theta$ and $\bf G$ are diagonal matrices. In such case the two responses are said to be {\em decoupled}.
To see further relations between the bivariate and the univariate models, we could try to solve a single univariate problem with the procedure in this paper. Suppose we want to solve for the best Box-Jenkins (1963) univariate dead band rule when the white noise is $\sigma_{\alpha}^2$, the IMA(1,1) parameter is $\theta$, the off target cost is $a$ and the adjustment cost $C$. Then we would set in our code $a_1=a_2=a$ and $\sigma_{1,\beta}^2=\sigma_{2,\beta}^2=(1-\theta)\sigma_{\alpha}^2$, apart from setting $\rho=0$. The solution thus obtained from minimizing $L'(\Lambda)$ will be related to the optimal solution found by Box and Jenkins, $\Lambda^{BJ}$, by the relation $\Lambda^{BJ}=\Lambda^*/\sqrt{2}$. The reason of this is the rotated nature of our dead area (Figure 1): $\Lambda^{BJ}$ is the width of the square but we are solving for $\Lambda^*$, half the length of the diagonal. We now illustrate the bivariate procedure with a practical example.\\

\noindent {\bf Example.-} As a practical application of the adjustment method developed and the optimal solutions obtained, consider a chemical mechanical planarization (CMP) process which is of critical importance in the manufacture of semiconductors. This is a polishing process in which there are typically two responses of interest (see, e.g., Moyne et al, 2000): the removal rate of silicon oxide (hereafter, $z_{t,1}$) which we suppose here to have a target equal to 2700, and the non-uniformity of the wafer (hereafter, $z_{t,2}$) with target equal to 500. Two controllable factors, down force ($X_{t,1}^{(0)}$) and table speed ($X_{t,2}^{(0)}$) can be adjusted to provide better control to target. The factors are in coded units. Here, the time index $t$ denotes the wafer number, assuming a single-wafer CMP machine is in use. To illustrate the methodology, we simulate this process from a somewhat modified model obtained from real experiments as reported in Del Castillo and Yeh (1998). Simulating the behavior of the process will allow us to see what would have occurred in the absence of any adjustments.

The model that is simulated for this illustration has a gain matrix equal to
\[
\bf G = \left(\begin{array}{cc}
547.6 & 616.3\\
-62.3 & -128.6
\end{array}
\right)
\]
and an IMA parameter matrix equal to
\[
\mathbf{\Theta} = \left(\begin{array}{cc}
0.4 & 0.1\\
0.3 & 0.5
\end{array}
\right)
\]
The covariance matrix of the bivariate normal white noise sequence is
\[
\bf C_{\alpha} =\left(\begin{array}{cc}
3600 & -1500\\
-1500 & 900
\end{array}
\right)
\]
In practice, estimates of the previous parameters could be obtained using multivariate time series techniques, see Reinsel (1993).

From the aforementioned data, we have that
\[
\mathbf{C}_{\beta} = (\mathbf{I}-\mathbf{\Theta}) \mathbf{C}_{\alpha} (\mathbf{I}-\mathbf{\Theta})'=\left(\begin{array}{cc}
1501 & -1268\\
-1268 & 1399
\end{array}
\right)
\]
thus $\rho=\rho_{\beta}=-0.8750$. Let us assume it costs $a_1 = 0.1$ dollars to have a removal rate
that deviates one unit (Amstrongs per time unit, in this case) from the desired target of 2700 during the processing of one wafer. Similarly, assume it costs $a_2=0.1$ dollars to have a wafer with a non-uniformity that deviates one unit (Amstrongs, in this case) from its desired target of 500. Assume the cost of making an adjustment in the ``recipe" used in processing each wafer equals $C=1000$ dollars, and includes the cost of re-starting the machine (sometimes test wafers are introduced after adjustments), machine downtime, and operator time. With the given cost structure and process information, we have that $C'=C/(a_1\sigma_{1,\beta}^2+a_2 \sigma_{2,\beta}^2)=3.448$. Minimizing $L'(\Lambda)$ with respect to $\Lambda$ we obtain the optimal limit $\Lambda^*=2.78$ with loss $L'(2.78)=1.3451$ and AAI$=E[N]=4.1589$ (wafers between adjustments), or approximately 24 adjustment will be made on average every 100 wafers are produced.\\

The resulting process adjustment procedure is as follows. A vector EWMA with parameter matrix $\bf L=I-\Theta$ provides one step ahead forecasts $\widehat z_{k+1,1}, \widehat z_{k+1,2}$ based on the measurements of the two responses. At each time instant $k$, the standardized bivariate series
$\mbox{\boldmath $U$}_k$ is computed as
\[
\mbox{\boldmath $U$}_k = \left(\begin{array}{c}
U_{k,1}\\
U_{k,2}
\end{array}
\right)
=
\left(\begin{array}{c}
\frac{\widehat z_{k+1,1}-X_{s,1}}{\sigma_{1,\beta}}\\
\frac{\widehat z_{k+1,2}-X_{s,2}}{\sigma_{2,\beta}}
\end{array}
\right)
\]
where $s <k $ is the last period an adjustment was made and where we use
$\mbox{\boldmath $X$}_k = \mathbf{G} \mbox{\boldmath $X$} _k^{(0)}$ with
$\mbox{\boldmath $X$}_k^{(0)}$ a vector containing the down force and table speed controllable factors as components. Whenever $|U_{k,1}+U_{k,2}|> 2.78 = \Lambda^*$ or $|U_{k,1}-U_{k,2}|>2.78$,
the controllable factors are changed such that $\mbox{\boldmath $X$}_{k+1} = \widehat{
\mbox{\boldmath $z$}}_{k+1}$, or, in terms of the original controllable factors, the new settings are
$\mbox{\boldmath $X$} _{k+1}^{(0)}=\mathbf{G}^{-1} \widehat{ \mbox{\boldmath $z$} }_{k+1}$.

Figure 3 shows the uncontrolled and controlled processes. Figure 4 shows the standardized quantities $U_{k,1}+U_{k,2}$ and $U_{k,1}-U_{k,2}$ on a ``adjustment chart" with limits at $\pm \Lambda^*
= \pm 2.78$. Finally, Figure 5 shows the corresponding values of the controllable factors. Horizontal segments imply no adjustments are made during such periods. In the particular simulation depicted, 30 adjustments were made. As it can be seen for the simulated data shown, the down force is reduced throughout the control session while the table speed was increased during the last few runs.

\begin{figure}[h]
\begin{center}
\resizebox{18cm}{!}{\includegraphics{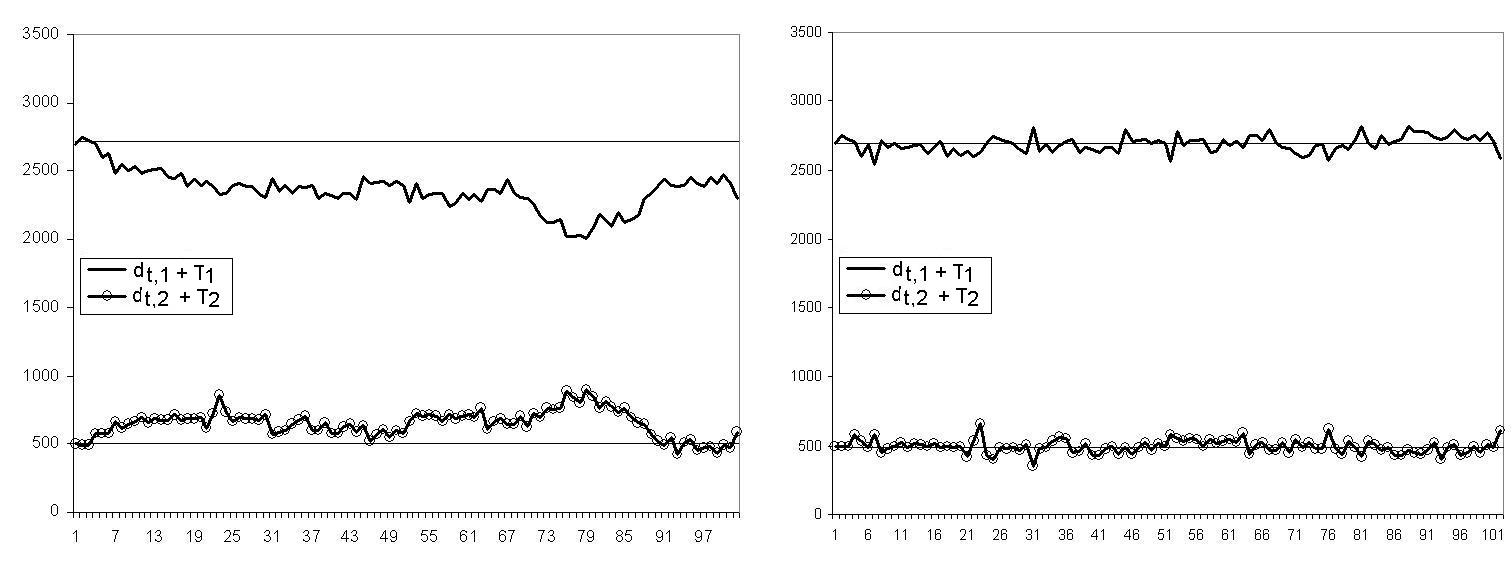}}
\caption{Realizations of the uncontrolled (left) vs. controlled (right) quality characteristics $(d_{t,l}+T_l, l=1,2)$ for the semiconductor example. Targets $T_l$ equal 2700 and 500 units for $l=1,2$, respectively.}
\rule{6cm}{0.3mm}
\end{center}
\end{figure}

\begin{figure}[h]
\begin{center}
\resizebox{12cm}{!}{\includegraphics{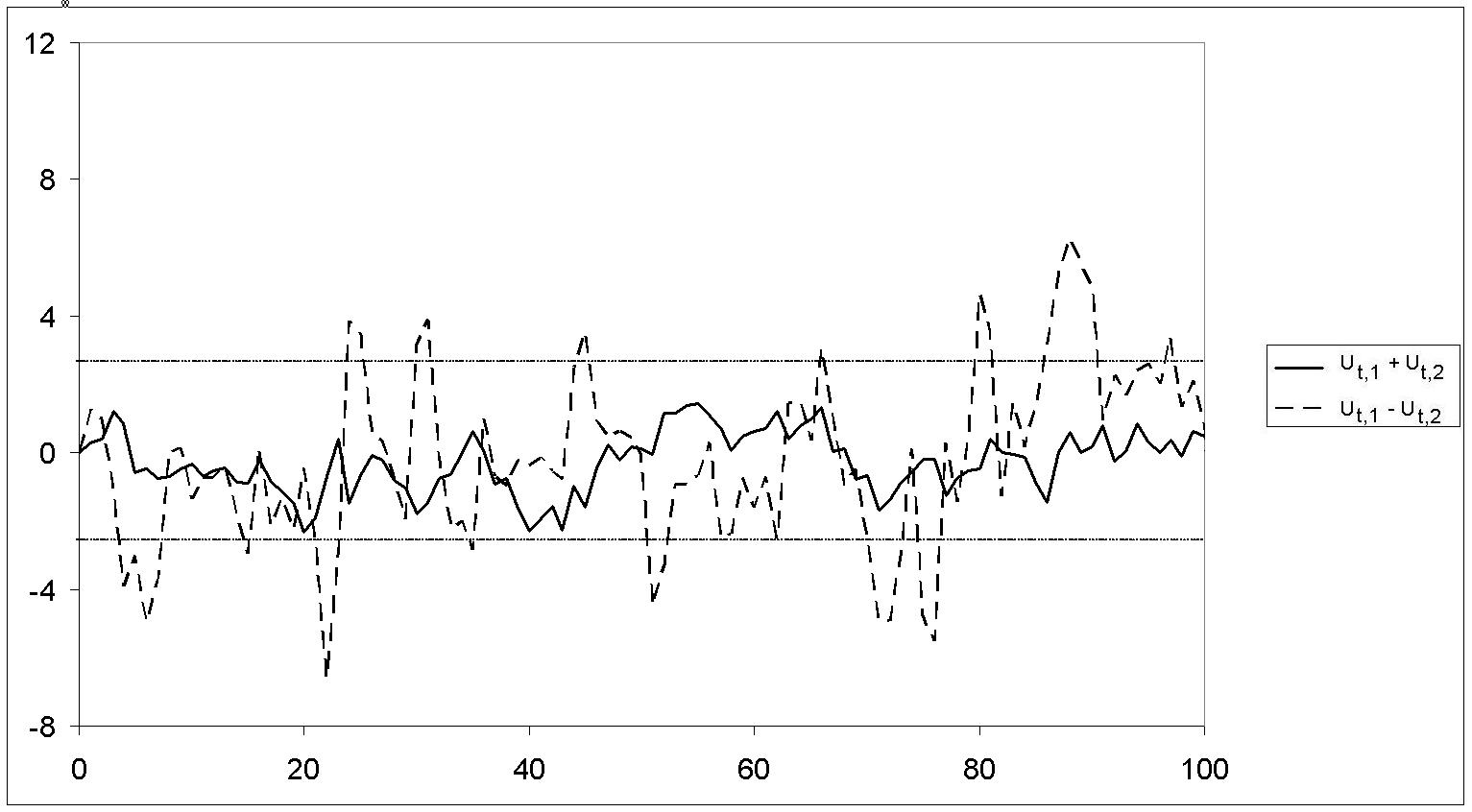}}
\caption{Standardized series $U_{t,1}+U_{t,2}$ and $U_{t,1}-U_{t,2}$ used to determine the time of the adjustments in the example. Adjustment limit $\Lambda^*=2.78$.}
\rule{6cm}{0.3mm}
\end{center}
\end{figure}

\begin{figure}[h]
\begin{center}
\resizebox{12cm}{!}{\includegraphics{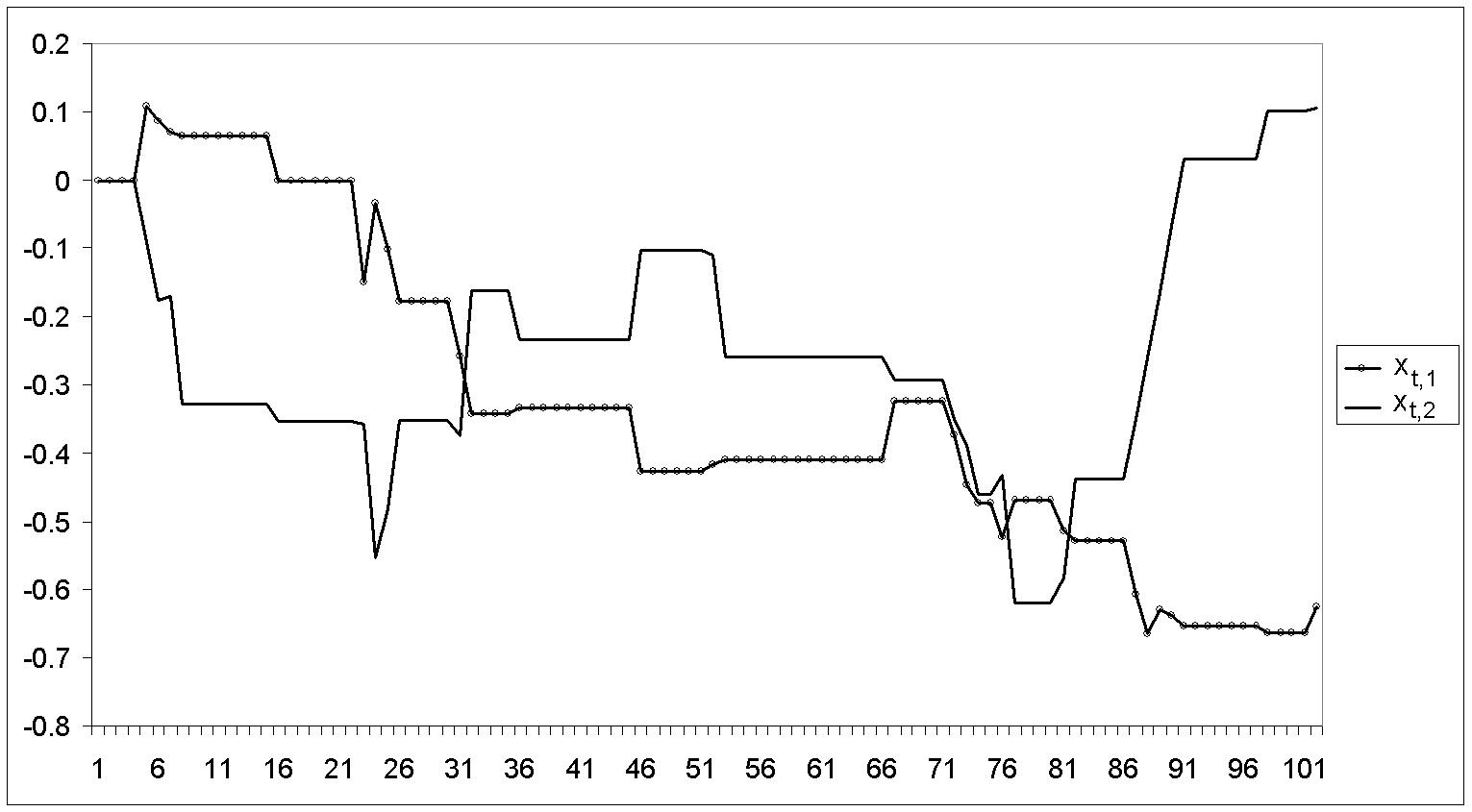}}
\caption{Levels for the controllable factors $X_{t,1}$ and $X_{t,2}$ for the example. Thirty adjustments were made.}
\rule{6cm}{0.3mm}
\end{center}
\end{figure}

\subsection*{References.}
{\AA}str\"om, K.J., (1970).
{\em Introduction to Stochastic Control Theory}. Academic Press, San Diego, CA.
\\[10pt]
Box, G.E.P., and Jenkins, G., (1963). ``Further Contributions
to Adaptive Quality Control: Simultaneous Estimation of
Dynamics: Nonzero Costs'',
{\em Bulletin of the International Statistical Institute}, 34, pp.\ 943-974.
\\[10pt]
Box G.E.P., G.M. Jenkins, and Reinsel, G. (1994). {\em Time Series
Analysis, Forecasting, and Control} 3rd.\ ed., Englewood Cliffs:
Prentice Hall.
\\[10pt]
Box, G.E.P., and Kramer, T., (1992).
``Statistical Process Monitoring and Feedback Adjustment--a Discussion'',
{\em Technometrics}, 34, 3, pp.\ 251-267.
\\[10pt]
Box, G.E.P., Luce\~{n}o, A. (1997). {\em Statistical Control by
Monitoring and Feedback Adjustment}. John Wiley \& Sons, New York,
NY.
\\[10pt]
Crowder, S.V., (1992). ``An SPC Model
for Short Production Runs: Minimizing Expected Cost,''
{\em Technometrics}, 34, pp.\ 64-73.
\\[10pt]
Del Castillo, E., (2002). {\em Statistical Process Adjustment for Quality Control}, New York: John Wiley \& Sons (Probability and Statistics Series).
\\[10pt]
Del Castillo, E. (2006), ``Statistical Process Adjustment: a brief
retrospective, current status and future research", {\em Statistica
Neerlandica}, 60(3), pp. 309-326.
\\[10pt]
Del Castillo, E., and Yeh, J.Y., (1998). ``An Adaptive Run-to-Run Optimizing Controller for Linear and Nonlinear Semiconductor Processes''
{\em IEEE Transactions on Semiconductor Manufacturing}, 11, 2, pp.\ 285-295.
\\[10pt]
Govind, N., Del Castillo, E., Runger, G., and Janakiram, M., (2018). ``Multivariate Bounded Adjustment Schemes", {\em Qual. Technology and Quant. Management},  15(2), pp. 253-273.
 \\[10pt]
Jensen, K.L., and Vardeman, S.B., (1993). ``Optimal Adjustment in
the Presence of Deterministic Process Drift and Random Adjustment
Error". {\em Technometrics}, 35, pp.\ 376-389.
\\[10pt]
Magni, L., Forgione, M., Toffanin, C., Dalla Man, C., Kovatchev, B., De Nicolao, G. and Cobelli, C., (2009). ``Run-to-run tuning of model predictive control for type 1 diabetes subjects: in silico trial". {\em J. of Diabetes Science and Technology}, 3 (5), pp. 1091-1098.
\\[10pt]
Moyne, J., Del Castillo, E., and Hurwitz, A., eds. (2000). {\em Run to run
control in semiconductor manufacturing}, CRC press,  Boca Raton, FL.
\\[10pt]
Liu, L., Ma, Y. and Tu, Y., (2013). ``Multivariate setup adjustment with fixed adjustment cost". {\em International Journal of Production Research}, 51(5), pp.1392-1404.
\\[10pt]
Reinsel, G.C., (1993). {\em
Elements of Multivariate Time Series Analysis}.
Springer-Verlag, New York, Berlin.
\\ [10pt]
Rogers, L.C.G., and Williams,  D. (1994) {\em Diffusions, Markov Processes,
and Martingales}. 2nd Edition. John Wiley \& Sons, Chichester, New York.
\\[10pt]
Ross, S.M., (1970). {\em Applied Probability Models with Optimization Applications}.
Holden-Day, San Francisco, Cambridge, London, Amsterdam.
\\[10pt]
Schmetterer, L., (1974). {\em Introduction to mathematical statistics} (Vol. 202). Springer Science \& Business Media.
\\[10pt]
Srivastava, M.\ S., and Wu, Y. (1991). ``A Second Order Approximation
to Taguchi's Online Control Procedure". {\em
Communications in Statistics -- Theory and Methods}, 20, 7, pp.\
2149-2168.
\\[10pt]
Woodall, W.H., and Del Castillo, E. (2014). ``An overview of George Box's contributions to process monitoring and feedback adjustment",   {\em Applied Stochastic Models in Business and Industry}, 30(1), pp. 53-61.

\begin{appendix}
\section{The Renewal Reward Process Property.}
\label {The Renewal Reward Process Property}
Consider an adjustment at time $s$. From the EWMA recursion (\ref {FOR_PROCESS_20})
for the one step ahead predictors and from the adjustment formula
$\; X_{s+1, l} = \hat{z}_{s+1, l}\;$ we obtain by induction
for $k=1,2,..$
\be \label {FOR_APPENDIX_10}
\wh { \mbox{\boldmath $d$}} _{s+k} \quad = _{ (\ref {FOR_CONTROL_10}) }
\quad   \hat{ \mbox{\boldmath $z$}}_{s+k} - \hat{
\mbox{\boldmath $z$}}_{s+1} \quad = \quad
{\bf L} \sum _{j=1} ^{k-1} \xalpha_{s+j}.
\ee
From equation (\ref {FOR_APPENDIX_10}) and from the alarm rule
(\ref {FOR_CONTROL_20}) it is obvious that the lengths
$N_1, N_2, N_3,...$ of periods between adjustments are independent and identically
distributed.
From equation
(\ref {FOR_PROCESS_10}) we obtain for $i=1,2,...$
$$ z_{s+i,l} \quad = \quad z_{s+1,l} \;+\;
\sum _{m=1} ^{i-1} ( \alpha_{s+m+1, l} - \theta_l
\alpha_{s+m, l} ) $$
and hence
\be \label {FOR_APPENDIX_20}
%z_{s+i,l} - R_{s+1,l} \quad = \quad
z_{s+i,l} - \hat{z}_{s+1,l}  \quad = _{ (\ref {FOR_PROCESS_30}) }
\quad   \alpha_{s+1, l}\;+\;
\sum _{m=1} ^{i-1} ( \alpha_{s+m+1, l} - \theta_l
\alpha_{s+m, l} ).\ee
From the assumptions on the white noise variables $\alpha_{t,l}$ and
from equation (\ref {FOR_APPENDIX_20}) it follows that
for $s_1 < s_2$, $1\leq i \leq s_2 -s_1$, $ j \geq 1$,
the differences $\; z_{s_1+i,l} - \hat{z}_{s+1,l}  \;$
and $\; z_{s_2+j,l} - \hat{z}_{s_2+1,l}  \;$ are independent and
normally distributed.
Taking into account that
 the adjustment formula is
$\; X_{s+1, l} = \hat{z}_{s+1, l}$, we can demonstrate that the vectors
%\be \label {FOR_CONTROL_40}
$\,
( {z}_{S_k+1, l} -X_{S_k+1, l},...,
{z}_{S_k+N_{k+1}, l} -X_{S_k+1, l})\,$ of the deviations from target,
indexed in $k\in \NAT$, are independent.
Hence by definition (\ref {FOR_LOSS_10}), the overall losses
$V_k$, $k=1,2,...$, in the periods between interventions are independent.
Since the lengths
$N_1, N_2, N_3,...$ of periods between adjustments are identically
distributed, the  losses
$V_k$, $k=1,2,...$, are also identically distributed.

\vspace{2mm}
Hence
the pairs $(N_k, V_k)$, $k=1,2,...$ are serially independent
and identically distributed, i.e., they constitute a
{\em renewal reward process}, see  Ross (1970).

\section {Integrals of the Normal Distribution Function.}
\label {Integrals of the Normal Distribution Function}
The incomplete gamma integral is defined by
\be \label {FOR_APPENDIX2_10}
\Gamma (y,a) \;\;=\;\;
\int _{a} ^{+\infty} u ^{y-1} \exp (-u) \, \mbox{d} u \quad
\mbox {for }\; a \in [0; +\infty).
\ee
For $a=0$ we obtain the customary gamma function $\; \Gamma(y) = \Gamma (y,0)$.
Integrals of the normal probability density function
$\; \varphi( y) = {1\over \sqrt {2\pi}} \exp \left (
{- y^2 \over 2} \right)\;$
can be expressed
by means of the incomplete gamma integral:
\be \label {FOR_APPENDIX2_20}
\int _{b} ^{+\infty} x ^j \varphi (x) \, \mbox{d} x \quad = \quad
{2 ^{ {j \over 2}-1 } \over \sqrt {\pi} } \; \Gamma \left ( {j+1\over 2},
{b^2 \over 2} \right) \qquad
\mbox {for }\; b \in [0; +\infty), \; j\in \NATO.
\ee
To prove formula (\ref {FOR_APPENDIX2_20}), substitute $\; u = {x^2\over 2}$.
%Observing
%$$ y^j  \Big [ 1- \phi (y)\Big] \;\;= \;\; \left
%( { \mbox{d} \over \mbox{d} y} {y^{j+1}  \over j+1}\right)
Integration by parts provides the formula
\be \label {FOR_APPENDIX2_30}
\int _{ A } ^{+\infty} y^j  \Big [ 1- \phi (y)\Big] \, \mbox{d} y \quad = \quad
{ -A^{j+1} \over j+1}
\Big[ 1 - \phi (A) \Big] \;+\; {1\over j+1}
\int _{A} ^{+\infty} y^ {j+1}   \varphi (y) \, \mbox{d} y \ee
for  $\; j\in \NATO$, $\; A\geq 0$,
and in particular
\be \label {FOR_APPENDIX2_40}
\int _{0} ^{+\infty} y^j  \Big [ 1- \phi (y)\Big] \, \mbox{d} y \quad = \quad
{1  \over j+1} \int _{0} ^{+\infty} y^ {j+1}   \varphi (y) \, \mbox{d} y \qquad
\mbox {for }\; j\in \NATO.
\ee
From formulae (\ref {FOR_APPENDIX2_20}), (\ref {FOR_APPENDIX2_30}),
(\ref {FOR_APPENDIX2_40}) we obtain
formulae for
integrals of the normal distribution function $\; \phi(x) \,= \,
\int _{-\infty}^x \varphi
(y)\, \mbox{d} y$.
%The values

\vspace{5mm}
\begin{bum}{Proposition} \label {PROPO_APPENDIX2_10}
For $\Delta \geq  0$, $q\in \NATO$, let
\be \label {FOR_APPENDIX2_50}
I_{1,1} (\Delta, q) \;\; = \;\; \int _{ \Delta} ^{+\infty}
x^q \Big[1- \phi (x-\Delta) \Big] \, \mbox{d} x,
\quad
I_{1,2} (\Delta, q) \;\; = \;\; \int _{ \Delta} ^{+\infty}
x^q \Big[1- \phi (x+\Delta) \Big] \, \mbox{d} x, \ee
\be \label {FOR_APPENDIX2_60}
I_{2,1} (\Delta, q) \;\; = \;\; \int _{ -\infty } ^{- \Delta}
x^q  \phi (x + \Delta)  \, \mbox{d} x,
\quad
I_ {2,2}  (\Delta, q) \;\; = \;\;
\int _{ -\infty } ^{- \Delta} x^q  \phi (x - \Delta)  \, \mbox{d} x
, \ee
\be \label {FOR_APPENDIX2_70}
%J (\Delta, q) \;\; = \;\; \int _ {-\Delta} ^{ \Delta}
%x^q \Big[ \phi (x+\Delta) - \phi (x-\Delta)\Big] \, \mbox{d} x,
%\quad
J_{0} (\Delta, q) \;\; = \;\; \int _0 ^{ 2\Delta}
(x-\Delta) ^q \phi (x) \, \mbox{d} x, \ee
and let $I(\Delta,q)$,
$J(\Delta,q)$ be defined by formulae (\ref {FOR_MOMENTSOF_I_INTEGRAL}),
(\ref {FOR_MOMENTSOF_J_INTEGRAL}).
Then we have:
\begin {description}
\item [(a)] $\displaystyle (-1)^q I_{2,1} (\Delta, q)
  \;\;= \;\;
I_{1,1} (\Delta, q) \;\; = \;\; \sum_{j=0} ^q
{q\choose j} { \Delta ^{q-j} \over j+1} \; \Gamma \left ( {j\over 2} +1 \right)
\; {2 ^ {j-1\over 2} \over \sqrt {\pi} } $.
\item [(b)] $\displaystyle (-1)^q
I_{2,2} (\Delta, q) \;\;= \;\;
I_{1,2} (\Delta, q) \;\; = \;\; $ \\ [2mm]
$\displaystyle \sum_{j=0} ^q
{q\choose j} { (- \Delta ) ^{q-j} \over j+1}
\left \{
- (2\Delta)^{j+1} \Big[ 1- \phi (2\Delta) \Big] \; +\;
\Gamma \left ( {j\over 2} +1, 2\Delta^2 \right)
{2 ^ {j-1\over 2} \over \sqrt {\pi} } \right \}$.
\item [(c)] $\displaystyle  J_{0} (\Delta, q)
  \;\;= \;\; $ \\[3mm]
$ \displaystyle {\Delta^{q+1} \over q+1} \left (
\phi (2\Delta) +  {(-1)^q \over 2} \right ) $ \\[3mm]
\mbox{  } \hfill \mbox{ } $\displaystyle \;-\; \;
{1\over q+1} \sum_{j=0} ^ {q+1} {q+1 \choose j}
(-\Delta)^{q+1-j}
{2 ^{ {j \over 2}-1 } \over \sqrt {\pi} } \; \left [
\Gamma \left ( {j+1\over 2} \right) -
\Gamma \left ( {j+1\over 2},
2 \Delta^2  \right) \right] $.
%%%%%%%%%%%%%%%%%%%%%%%%%%%%%%%%%%%%%%%%%%%
\item [(d)]
$\displaystyle I (\Delta, q) \;\; = \;\; \Big [ 1+ (-1)^q \Big ]
\Big[ I_{1,1} (\Delta,q) \, -\, I_{1,2} (\Delta,q) \Big]$, where
$I(\Delta,q)$ is defined by formula (\ref {FOR_MOMENTSOF_I_INTEGRAL}).

\item [(e)]
$ \displaystyle J (\Delta, q) \;\; = \;\;
\Big [ 1+ (-1)^q \Big ] \left [ J_0 (\Delta, q) \;-\; {\Delta^{q+1} \over q+1} \right]$,
 where
$J(\Delta,q)$ is defined by formula (\ref {FOR_MOMENTSOF_J_INTEGRAL}).
\end {description}
\end{bum}

\vspace{1mm}
\noindent
{\em Proof of assertion (a) of proposition
\ref {Integrals of the Normal Distribution Function}.\ref {PROPO_APPENDIX2_10}}.
Substituting $\; y =-x\;$ we obtain
%
%\noindent
%\vspace{2mm}
%$\displaystyle
$$ (-1)^q
\int _{ -\infty } ^{- \Delta}
x^q  \phi (x+\Delta)  \, \mbox{d} x \quad = \quad
- \int _{ -\infty } ^{- \Delta}
(-1) (-x)^q  \Big [ 1- \phi (-x-\Delta)  \Big] \, \mbox{d} x \quad
= \quad  I_{1,1} (\Delta, q) .$$
%
%\noindent
%\vspace{2mm}
Substituting $\; z = y-\Delta\;$ we obtain
%\noindent
%\vspace{2mm}
%$\displaystyle
$$ \begin {array} {l}
\displaystyle  I_ {1,1} (\Delta, q) \quad = \quad
\int _{ 0} ^{+\infty}
(z+\Delta) ^q \Big[1- \phi (z) \Big] \, \mbox{d} z \quad =
_{ (\ref {FOR_APPENDIX2_40}) } \quad \\[4mm]
\displaystyle  \sum _{ j=0}^q {q\choose j} { \Delta^ {q-j} \over j+1}
\int _{0} ^{+\infty} y^ {j+1}   \varphi (y) \, \mbox{d} y \quad
=_{ (\ref {FOR_APPENDIX2_20}) }
\sum _{ j=0}^q {q\choose j} { \Delta^ {q-j} \over j+1 }
\; \Gamma \left ( {j\over 2} +1 \right)
\; {2 ^ {j-1\over 2} \over \sqrt {\pi} } .
\end {array}
$$
Assertion (b) of proposition
\ref {Integrals of the Normal Distribution Function}.\ref {PROPO_APPENDIX2_10}
is proved analogously: the first identity is obtained by
substituting $\; y=-x$; the second identity is obtained by substituting $\;
z= y+\Delta$, and then
using formulae  (\ref {FOR_APPENDIX2_20}) and
(\ref {FOR_APPENDIX2_30}).
%\mbox{ } \hfill $ \bullet$

\vspace{2mm}
\noindent
{\em Proof of assertion (c) of proposition
\ref {Integrals of the Normal Distribution Function}.\ref {PROPO_APPENDIX2_10}}.
Integration by parts provides
$$ J_{0} (\Delta, q)
  \;\;= \;\; \left [ { ( x-\Delta)^{q+1} \over q+1 } \phi(x) \right] \bigg \vert
_{0} ^{2\Delta} \;-\; {1\over q+1} \int_0 ^{2\Delta} (x-\Delta)^{q+1}
\varphi (x) \, \mbox{d} x \quad = \quad $$
$$
 {\Delta^{q+1} \over q+1} \left (
\phi (2\Delta) +  {(-1)^q \over 2} \right ) \;-\; \;
{1\over q+1} \sum_{j=0} ^ {q+1} {q+1 \choose j}
(-\Delta)^{q+1-j} \int_0 ^{2\Delta} x^{ j }
\varphi (x) \, \mbox{d} x .$$
By formulae  (\ref  {FOR_APPENDIX2_10}) and (\ref  {FOR_APPENDIX2_20})
we obtain the assertion on $J_0 (\Delta, q)$.

\vspace{2mm}
\noindent
{\em Proof of assertion (d) of proposition
\ref {Integrals of the Normal Distribution Function}.\ref {PROPO_APPENDIX2_10}}.
The symmetry relation $\; \phi (-y) = 1- \phi (y)\;$ for the distribution function
of the standard normal distribution provides
$$ I(\Delta, q) \quad = _{ (\ref {FOR_MOMENTSOF_I_INTEGRAL}), \;
(\ref {FOR_APPENDIX2_50}), \; (\ref {FOR_APPENDIX2_60}) } \quad
I_{1,1} (\Delta, q) - I_{1,2} (\Delta, q) + I_{2,1} (\Delta, q)
- I_{2,2} (\Delta, q)
\quad = _{ (a), (b) } \quad $$
$$
\Big [ 1+ (-1)^q \Big ]
\Big[ I_{1,1} (\Delta,q) \, -\, I_{1,2} (\Delta,q) \Big].$$

\vspace{2mm}
\noindent
{\em Proof of assertion (e) of proposition
\ref {Integrals of the Normal Distribution Function}.\ref {PROPO_APPENDIX2_10}}.
The substitution $\; u=-z\;$ provides
$$ \int_{-2\Delta} ^0 (z+\Delta)^{q}
\phi (z) \, \mbox{d} z \quad = \quad
(-1)^q \int_0 ^{2\Delta}  (u-\Delta)^{q}
\Big[ 1-\phi (u)\Big] \, \mbox{d} u \quad = \quad $$
$$
(-1)^q \left \{ { \Delta^{q+1} \over q+1} \Big( 1 - (-1)^ {q+1} \Big)
\;-\; J_0 (\Delta, q) \right\}.$$
Using this result and
substituting $\, y= x+\Delta\,$ and, respectively,
$\, z = x-\Delta\,$ in the definition
of $J(\Delta, q)$ in formula
(\ref {FOR_APPENDIX2_70}),  we obtain
$$ J (\Delta, q) \quad = \quad
\int_0 ^{2\Delta} (y-\Delta)^{q}
\phi (y) \, \mbox{d} y \;-\;
\int_{-2\Delta} ^0 (z+\Delta)^{q}
\phi (z) \, \mbox{d} z
\quad = \quad$$
$$
J_0 (\Delta, q) \;-\; (-1)^q \left \{ { \Delta^{q+1} \over q+1} \Big( 1 - (-1)^{q+1} \Big)
\;-\; J_0 (\Delta, q) \right\} \quad = \quad $$
$$ \Big( 1+ (-1)^q\Big) \left \{ J_0 (\Delta, q) \; -
\; {  \Delta^{q+1} \over q+1} \right\}.$$
\mbox{ } \hfill $ \bullet$

\section {Three Martingales.}
\label {Some Useful Martingales}
Let the family $\Big( (u_{n,1}, u_{n,2}) \Big)_{\NAT}$ of variables
introduced in Section
\ref {The Standardized Loss Function}
be adapted to its  natural filtration $( {\cal A}_n)_ {\NAT}$, i.e.,
let $( {\cal A}_n)_ {\NAT}$ be the sequence of smallest
$\sigma$-algebras
with ${\cal A}_1 \subset {\cal A}_2 \subset ...$ where
$(u_{n,1}, u_{n,2})$ is Borel-measurable with respect to
${\cal A}_n$ for $n \in \NAT$.
Then $\; U_{n,l} = u_{1,l} +...+ u_{n,l}\;$ is Borel-measurable with
respect to ${\cal A}_n$ for $n \in \NAT$, $l=1,2$.
The martingale property with respect to the filtration
$( {\cal A}_n)_ {\NAT}$ is determined by the
conditional expectations
$\; E[ \, \cdot\, \vert {\cal A}_n] =
E[ \, \cdot\, \vert u_{n,1}, u_{n,2},..., u_{1,1}, u_{1,2} ] $.
In this sense, the following sequences $(R_{n,l})_{n\in \NAT}$,
$(Y_{n,l})_{n\in \NAT}$, $(Z_{n,l})_{n\in \NAT}$ with
\ba \label {FOR_MARTINGALE_10}
R_{n,l} \;\;= \;\; U_{n,l} ^2 -n, \qquad
Y_{n,l} \;\;= \;\; \sum _{ k=1} ^ {n-1} U_{k,l} ^2 -
n U_{n,l}^2 + {n (n+1) \over 2}, \quad
\ea
\ba \label {FOR_MARTINGALE_20}
Z_{n,l} \;\;= \;\; {1\over 6}  U_{n,l} ^4 -
n U_{n,l}^2 + {n^2 \over 2},
\ea
are martingales.
For the proof, we observe that
\ba \label {FOR_MARTINGALE_30}
E[ u_{k+1,l} ^q U_{k,l}^r \vert {\cal A}_k] \quad
=  \quad
U_{k,l}^r E[ u_{k+1,l} ^q ] \quad = \quad
U_{k,l}^r \cdot \ZWEI
{ 0, } { \mbox {if $q$ is odd,} }
{ { q! \over \left ({q\over 2}\right) ! 2 ^{q/2} },}
{ \mbox {if $q$ is even,}}
\ea
since $U_{k,l}^r$ is measurable with respect to ${\cal A}_k$ and
$u_{k+1,l} ^q $ is independent of ${\cal A}_k$.
Hence
$$ E[ R_{n+1,l} \vert {\cal A}_n] \quad
=  \quad
E[ R_{n,l}  +2  u_{n+1,l}  U_{n,l} +  u_{n+1,l}^2-1\vert {\cal A}_n ] \quad
=_{ (\ref {FOR_MARTINGALE_30}) }  \quad R_{n,l},$$
$$ E[ Y_{n+1,l} \vert {\cal A}_n] \quad
=  \quad
E[ Y_{n,l}  - (n+1)  u_{n+1,l}^2 - 2  u_{n+1,l} U_{n,l} +  n+1 \vert {\cal A}_n] \quad
=_{ (\ref {FOR_MARTINGALE_30}) }   \quad Y_{n,l},$$
\begin{eqnarray}
\lefteqn{ Z_{n+1,l}    \quad = \quad } \nonumber \\[2mm]
&&
Z_{n,l}  +
  {2\over 3}  u_{n+1,l} U_{n,l}^3 + u_{n+1,l}^2 U_{n,l}^2
              {2\over 3}  u_{n+1,l}^3  U_{n,l} + \nonumber \\[2mm]
 && {1\over 6} u_{n+1,l}^3
-U_{n,l}^2 - 2(n+1) u_{n+1,l} U_{n,l} - (n+1) u_{n+1,l}^2 +
n +{1\over  2}
\end {eqnarray}
and hence by (\ref {FOR_MARTINGALE_30})
$\; E[ Z_{n+1,l} \vert {\cal A}_n] = Z_{n,l}$.

\end {appendix}

\end{document}